\documentclass[journal]{IEEEtran}

\ifCLASSINFOpdf

\else

\fi

\usepackage{mathrsfs}
\usepackage{amsmath}
\usepackage{amsfonts}
\usepackage{amsthm}
\usepackage{amssymb}
\usepackage{graphicx}
\usepackage{subfigure}
\usepackage{indentfirst}
\usepackage{array}
\usepackage{epstopdf}
\usepackage{cite}
\usepackage{enumerate}
\usepackage{bm}
\usepackage{dsfont}
\usepackage[linesnumbered,ruled,vlined]{algorithm2e}
\usepackage{multirow}
\usepackage{verbatim}
\usepackage{stfloats}
\usepackage{makecell}
\usepackage{cancel}
\usepackage{threeparttable}
\usepackage[dvipsnames]{xcolor}
\usepackage{pythonhighlight}
\usepackage[frozencache,cachedir=minted-cache]{minted}
\usepackage[colorlinks, linkcolor=blue]{hyperref}

\usepackage{listings}
\lstdefinestyle{Python}{
	language        = Python,
	basicstyle      = \ttfamily,
	keywordstyle    = \color{blue},
	keywordstyle    = [2] \color{teal}, 
	stringstyle     = \color{green},
	commentstyle    = \color{red}\ttfamily
}

\definecolor{lightgray}{rgb}{0.83, 0.83, 0.83}

\definecolor{revised}{HTML}{b71a3b}

\SetKwComment{Comment}{//}{}
\SetKwBlock{Begin}{Function}{end}

\begin{document}
	\lstset{
		frame       = single,
		numbers     = left,
		showspaces  = false,
		showstringspaces    = false,
		captionpos  = t,
		caption     = \lstname
	}

\title{Improved Nonlinear Transform Source-Channel Coding to Catalyze Semantic Communications}

\author{Sixian Wang,~\IEEEmembership{Graduate Student Member,~IEEE},
        Jincheng Dai,~\IEEEmembership{Member, IEEE},
        Xiaoqi Qin,~\IEEEmembership{Member, IEEE},
        Zhongwei Si,~\IEEEmembership{Member, IEEE},
        Kai Niu,~\IEEEmembership{Member, IEEE},
        and Ping Zhang,~\IEEEmembership{Fellow, IEEE}

\thanks{This work was supported in part by the National Natural Science Foundation of China under Grant 62293481, Grant 92067202, Grant 62001049, and Grant 62071058, the BUPT Excellent Ph.D. Students Foundation under Grant CX2022146, the Fundamental Research Funds for the Central Universities, and the Natural Science Foundation of Beijing Municipality under Grant 4222012. \emph{Corresponding author: Jincheng Dai.}}

\thanks{Sixian Wang, Jincheng Dai, Zhongwei Si, and Kai Niu are with the Key Laboratory of Universal Wireless Communications, Ministry of Education, Beijing University of Posts and Telecommunications, Beijing 100876, China (e-mail: daijincheng@bupt.edu.cn).}

\thanks{Xiaoqi Qin and Ping Zhang are with the State Key Laboratory of Networking and Switching Technology, Beijing University of Posts and Telecommunications, Beijing 100876, China.}

\vspace{-1em}

}

\maketitle

\begin{abstract}
Recent deep learning methods have led to increased interest in solving high-efficiency end-to-end transmission problems. These methods, we call \emph{nonlinear transform source-channel coding (NTSCC)}, extract the semantic latent features of source signal, and learn entropy model to guide the joint source-channel coding with variable rate to transmit latent features over wireless channels. In this paper, we propose a comprehensive framework for improving NTSCC, thereby higher system coding gain, better model compatibility, more flexible adaptation strategy aligned with semantic guidance are all achieved. This new sophisticated NTSCC model is now ready to support large-size data interaction in emerging XR, which catalyzes the application of semantic communications. Specifically, we propose three useful improvement approaches. First, we introduce a contextual entropy model to better capture the spatial correlations among the semantic latent features, thereby more accurate rate allocation and contextual joint source-channel coding method are developed accordingly to enable higher coding gain. On that basis, we further propose a response network architecture to formulate \emph{compatible} NTSCC, i.e., once-learned model supports various bandwidth ratios and channel states that benefits practical deployment greatly. Following this, we propose an online latent feature editing mechanism to enable more flexible coding rate allocation aligned with some specific semantic guidance. By comprehensively applying the above three improvement methods for NTSCC, a deployment-friendly semantic coded transmission system stands out finally. Our improved NTSCC system has been experimentally verified to achieve a better rate-distortion efficiency versus the state-of-the-art engineered VTM + 5G LDPC coded transmission system with lower processing latency. Our code is available at  \href{https://semcomm.github.io/ntscc\_plus/}{https://semcomm.github.io/ntscc\_plus/}.
\end{abstract}


\IEEEpeerreviewmaketitle

\section{Introduction}\label{section_introduction}

\subsection{Related Work and Motivation}

\IEEEPARstart{I}{n} recent years, the rapid emergence and prosperity of immersive applications, such as ultra-large scale image/video interaction in extended reality (XR) \cite{ling2022future,xu360,xie2022neural}, will have higher demand on end-to-end communications. They are raising the challenges to the traditional separate development routes of source compression and channel transmission domains. In the face of very big data transmission and massive emerging applications, on the one hand, current wireless transmission systems suffer from time-varying channel conditions, in which case the mismatch between communication rate and channel capacity results in obvious \emph{cliff-effect}, i.e., the performance breaks down when the channel capacity goes below communication rate. On the other hand, the widely-used entropy coding, which compresses the source representation into bit sequences, is quite sensitive to the channel decoded results, any left bit error can lead to catastrophic error propagation in entropy decoding. Moreover, the original source compression paradigm based on pixel/voxel signal processing cannot align with the requirements of low-level human perception and high-level semantic analysis. It is the very time to develop new coded transmission paradigms to support real-time broadband communications (RTBC) toward emerging media end-to-end transmission tasks.

In this context, \emph{semantic coded transmission (SCT)} \cite{gunduz2022beyond,dai2021semantic,zhang2021toward}, has emerged to bridge an extent separate research domains of representation learning, source compression, and channel transmission, and attempts to optimize compactness, efficiency, and robustness jointly from a unified perspective of end-to-end rate-distortion (RD) tradeoff. Particularly, the initial line of work \cite{djscc,djsccl,saidutta2021joint} using deep neural networks (DNNs) to realize joint source-channel coding (deep JSCC) has led to steady progress of end-to-end transmission performance over past years, reaching or even surpassing state-of-the-art (SOTA) source codecs combined with channel codecs on small-size images (e.g., CIFAR dataset, $32\times 32$ images). Later, inspired by the landmark work of Ball{\'e} \emph{et al.} on nonlinear transform coding (NTC) based source compression \cite{balle2020nonlinear,balle2016,balle2018}, nonlinear transform source-channel coding (NTSCC) \cite{dai2022nonlinear} was proposed as one of the technical foundations of SCT. Codecs based on NTSCC divide the task of end-to-end data transmission into three modularized components: transform, entropy modeling, and JSCC. All three components are formulated by deep neural networks using its nonlinear processing property: autoencoder-like networks are adopted as flexible nonlinear transforms to extract the semantic latent features of the source signal, deep generative models are leveraged as powerful learnable entropy models performed on the latent features, and powerful learned JSCC codecs with dynamic rates are used to transport latent features. Since the entropy model will guide to allocate more channel bandwidth in JSCC codec to the more important contents which are crucial for the reconstruction of the source signal. This kind of content-adaptive mechanism enables the superior system coding gain over the plain deep JSCC. In addition, NTSCC has shown on-par or better transmission RD performance than SOTA engineered source and channel codecs in high-resolution image/video sources (e.g., CLIC21 dataset, 2K images \cite{CLIC21}, etc.) \cite{wang2023wireless}.

Though it seems promising, there are still many problems to solve before those NTSCC models can be deployed in practice, which are analyzed as the following aspects.

\subsubsection{Correlation Modeling}

Spatial and temporal redundancy is an important basis of image and video compression, which is also utilized in NTSCC. Strong self-correlation exists in the raw source data, for example, adjacent image pixels are likely to have a stronger causal relationship. Empirically, the more abstract semantic latent features output by nonlinear analysis transform still keep such redundancy because of locality, even when using means-scale Gaussian hyperprior to obtain more accurate entropy model in the original NTSCC.

Motivated by the success of autoregressive priors in probabilistic generative models \cite{minnen2018}, it is expected to introduce \emph{context} model to predict the probability of unknown latent features conditioned on features that have been already decoded. These context models are more powerful to capture the correlations in the semantic latent features, which can guide more accurate rate allocation in variable-rate JSCC. Accordingly, a contextual JSCC codec is also needed to leverage the explicitly modeled correlations among latents to improve performance.

\subsubsection{Model Compatibility}

Given a model that can provide an accurate entropy estimate of a latent feature map, the previous NTSCC frameworks optimize their networks by minimizing the weighted sum of the RD pairs by using the method of Lagrange multipliers. The Lagrange multiplier $\lambda$ introduced in the Lagrangian is treated as a \emph{hyperparameter} to train a network for a desired tradeoff between the channel bandwidth cost and the quality of transmitted source signal. In addition, the scaling factor $\eta$ that controls the relation between the estimated entropy of each latent feature and its JSCC code length is another term to affect the NTSCC model performance. This implies that one needs to train and deploy various separate NTSCC models for rate adaptation. One way is to re-train a network while varying $\lambda$ or $\eta$. However, this is impractical when we operate at a broad range of the RD curve and the size of networks is large. It is necessary to develop \emph{compatible} NTSCC to support a large range of rate adjustments with only one well trained model, which can greatly facilitate the practical deployment greatly. Moreover, once-trained NTSCC model needs to be adapted to various channel conditions.

\subsubsection{Model Adaptability}

Although the existing optimized NTSCC models have proven to be very successful in minimizing the end-to-end expected RD cost over a full source dataset and ergodic wireless channel responses, they are yet unlikely to be optimal for every test instance due to the limited model capacity. Existing models only pay attention to an average low RD cost on the training set. For a given input data sample and wireless channel response, such a learned codec may not effectively capture the data's semantic features and the channel state at that specific instance, resulting in suboptimal transform and coding during the model inference stage. Moreover, once trained model targeted to a given RD metric cannot be smoothly adapted to other RD targets when incorporating some additional semantic guidance, e.g., human perceptual metric or region-of-interest (ROI) guidance. Consequently, it is essential to develop a mechanism for online editing of semantic latent features that can adapt to a variety of RD targets and semantic guidance requirements.

\subsection{Summary of Contributions}

To tackle the aforementioned problems, we propose a series of improvement methods to formulate a deployment-friendly NTSCC system to catalyze semantic communications. Specifically, the contributions are summarized as follows.

\subsubsection{Contextual Modeling for Higher Coding Gain}

We introduce the novel checkerboard context model (CCM) \cite{he2021checkerboard} to better exploit the correlative probabilistic structure in the latent features. It is parallelization-friendly thus avoiding the horribly low computational efficiency in the traditional autoregressive context modeling \cite{minnen2018}. Accordingly, we develop the contextual JSCC codec architecture to collaborate with the two-stage CCM. Our \emph{contextual NTSCC} achieves clearly better end-to-end RD performance with good computational efficiency.

\subsubsection{Response Network Enabled NTSCC Framework for Better Compatibility}

We propose a response network architecture to build a computationally efficient \emph{compatible NTSCC} framework for learning optimal parameters corresponding to various hyperparameters in a single training run. By this means, once-trained NTSCC model can support a very wide range of RD performance and different wireless channel conditions during deployment. The key idea is to explicitly build response functions that embed the rate-control hyperparameter $\lambda$ to the optimal parameters of the semantic analysis/synthesis transform modules, and embed the channel state information to the optimal variable-rate JSCC codec parameters directly. We justify that under various channel states, our compatible NTSCC architecture can construct the end-to-end RD curve without additional training and can be deployed with significantly less hyperparameter tuning.

\subsubsection{Online Latent Feature and JSCC Codec Editing Mechanism for More Flexible Adaptability}

Based on the above two improvements, we further propose an online semantic feature editing and JSCC codec updating mechanism. It indeed leverages the deep learning model's overfitting property, thereby the latent feature and the JSCC codec can for instance be updated after deployment, which further lead to substantial gains in terms of any target RD performance. It is so flexible that we do not even need to know what the exact RD target functions are during training. Therefore, a flexible SCT framework is established to accommodate new introduced semantic guidance in XR, e.g., ROI map, human perceptual metric, etc.

By comprehensively applying the above three improvement methods for NTSCC, a deployment-friendly semantic coded transmission system stands out finally. Extensive experimental results verify the effectiveness and efficiency of our proposed methods. Through these experiments, we show the potential of \emph{improved NTSCC} supporting the communication requirements of the emerging XR.

The remainder of this paper is organized as follows. In Section \ref{section_preliminaries}, we review the architecture and properties of NTSCC based semantic communication system. In Section \ref{section_context}, we present the CCM and its corresponding contextual JSCC codec architecture. In Section \ref{section_versatile}, we present some important insights about how the hyperparameter $\lambda$ and the channel state change the NTSCC model parameters, then the response networks to realize the bandwidth compatible NTSCC are given. On that basis, Section \ref{section_overfit} presents the flexible NTSCC architecture via the online semantic latent feature editing and JSCC codec updating. Section \ref{section_results} shows experimental results to quantify our performance gain, and some valuable discussions are also given. Finally, Section \ref{section_conclusion} concludes this paper.

\emph{Notational Conventions:} Throughout this paper, lowercase letters (e.g., $x$) denote scalars, bold lowercase letters (e.g., $\boldsymbol{x}$) denote vectors. In some cases, $x_i$ denotes the elements of $\boldsymbol{x}$, which may also represent a subvector of $\boldsymbol{x}$ as described in the context. Bold uppercase letters (e.g., $\boldsymbol{X}$) denote matrices, and $\boldsymbol{I}_m$ denotes an $m$-dimensional identity matrix. $\ln (\cdot)$ denotes the natural logarithm, and $\log (\cdot)$ denotes the logarithm to base $2$. $p_x$ denotes a probability density function (pdf) with respect to the random variable $x$. In addition, $\mathbb{E} (\cdot)$ denotes the statistical expectation operation, and $\mathbb{R}$ denotes the real number set. Finally, $\mathcal{N}(x|\mu, \sigma^2) \triangleq (2\pi \sigma^2)^{-1/2} \exp(-(x - \mu)^2/(2\sigma^2))$ denotes a Gaussian function, and $\mathcal{U}(a-u,a+u)$ stands for a uniform distribution centered on $a$ with the range from $a-u$ to $a+u$.

\section{NTSCC based Semantic Communication System}\label{section_preliminaries}

\begin{figure*}[t]
\setlength{\abovecaptionskip}{0.cm}
\setlength{\belowcaptionskip}{-0.cm}
	\centering{\includegraphics[scale=0.325]{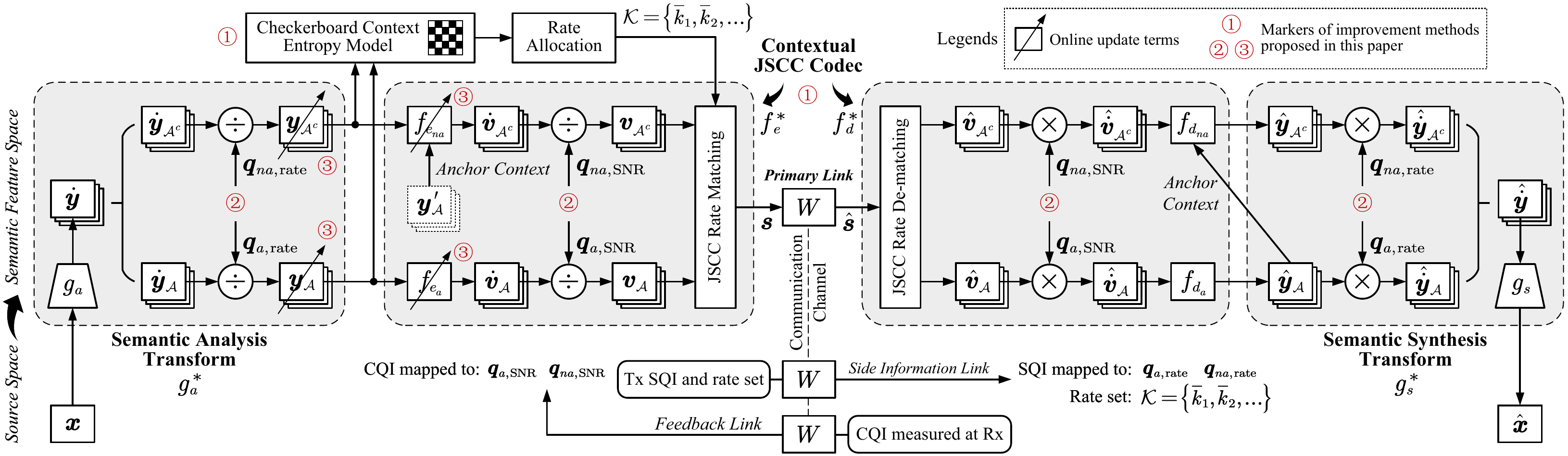}}
	\caption{The whole NTSCC++ system architecture for semantic communications, which includes the semantic analysis transform $g_a^*$, the contextual JSCC encoder $f_e^*$, the contextual JSCC decoder $f_d^*$, the semantic synthesis transform $g_s^*$, and the checkerboard context entropy model. Compared to the original NTSCC system, three improvement methods are marked with circled numbers. \textcircled{\scriptsize{1}} denotes the contextual modeling and contextual JSCC toward higher system coding gain (NTSCC+); \textcircled{\scriptsize{2}} denotes the rate compatible and channel adaptive strategy to enable compatible NTSCC+; \textcircled{\scriptsize{3}} denotes the online latent feature and JSCC codec editing methods to enable more flexible adaptability for NTSCC++.}\label{Fig_ntsccpp_system_architecture}
\end{figure*}

The idea of NTSCC stems from the landmark work of Ball\'{e} \emph{et al.} on nonlinear transform coding (NTC) \cite{balle2020nonlinear,balle2016,balle2018}. Given the pristine data sample $\boldsymbol{x}$, e.g., an image $\boldsymbol{x}$ modeled as a vector of pixel intensities $\boldsymbol x \in {\mathbb R}^m$, it is first transformed into semantic latent representation $\boldsymbol{y}$ using a DNN-based nonlinear analysis transform $g_a$. In data compression tasks, $\boldsymbol{y}$ will be quantized as discrete-valued latent representation $\boldsymbol{\bar y}$, followed by entropy encoding \cite{witten1987arithmetic} to convert $\boldsymbol{\bar y}$ into bit sequence. This bit sequence will be fed into entropy decoding to losslessly reconstruct $\boldsymbol{\bar y}$, and another nonlinear synthesis transform DNN module $g_s$ uses $\boldsymbol{\bar y}$ to reconstruct the decoded data $\boldsymbol{\hat x}$. The $g_a$ and $g_s$ are jointly optimized under the RD constraint. In wireless communication systems, the above source compressive coding paradigm relies heavily on advanced channel coding and signal processing techniques to ensure the transmitted bit sequence to be losslessly recovered. This separation-based approach has been employed in many current communication systems, as the binary representations of various source data can be seamlessly transmitted over arbitrary wireless channels by changing the underlying channel code.

However, with increasing demands on low-latency wireless data delivery applications such as extended reality (XR), the limits of the separation-based design begin to emerge. Current wireless data transmission systems suffer from time-varying channel conditions, in which case the separation-based design leads to significant \emph{cliff-effect} when the channel condition is below the level anticipated by the channel code \cite{djscc}. Furthermore, the widely-used entropy coding is quite sensitive to the variational estimate of the marginal distribution of the source latent representation. Small perturbations on this marginal can lead to the catastrophic error propagation in entropy decoding \cite{rissanen1981universal}. As a result, this non-determinism issue in transmitter vs. receiver will lead to severe performance degradation.

To address the above issues, the idea in NTSCC \cite{dai2022nonlinear} is to \emph{replace entropy coding and channel coding} with a trained DNN for joint source channel coding. In particular, a variable-rate JSCC is proposed to transmit the nonlinear transform code $\boldsymbol{y}$ directly. This approach results in better end-to-end transmission performance and maintains system robustness to unpredictive wireless channels. The whole procedure of NTSCC is depicted in \eqref{eq_trans_process}, where the latent code $\boldsymbol{y}$ is fed into both the hyper-prior path ($h_a$ and $h_s$) and the variable-rate JSCC encoder $f_e$. On the one hand, $h_a$ summarizes the distribution of mean values and standard deviations of $\boldsymbol{y}$ in the \emph{hyperprior} $\boldsymbol{z}$. The transmitter utilizes $\boldsymbol{z}$ to estimate both the mean vector $\boldsymbol{\mu}$ and the standard deviation vector $\boldsymbol{\sigma}$, which will be further used to determine the required bandwidth for transmitting the latent representation. On the other hand, $f_e$ encodes $\boldsymbol{y}$ as the channel-input sequence $\boldsymbol{s} \in \mathbb{R}^k$, and the received sequence is ${\boldsymbol{\hat s}} = W( \boldsymbol{s} )$, with a transition probability of ${{p_{{\boldsymbol{\hat s}}| {\boldsymbol{s}} }}( {{\boldsymbol{\hat s}}| \boldsymbol{s} } )}$. In this paper, we consider the general fading channel model such that the transfer function is ${\boldsymbol{\hat s}} = W( \boldsymbol{s}|\boldsymbol{h} ) = \boldsymbol{h} \odot \boldsymbol{s} + \boldsymbol{n}$ where $\odot$ is the element-wise product, $\boldsymbol{h}$ is the channel gain vector, and each component of the noise $\boldsymbol{n}$ is independently sampled from a Gaussian distribution, i.e., $\boldsymbol{n} \sim p_{\boldsymbol{n}} \triangleq \mathcal{N}(\boldsymbol{0}, {\sigma_n^2}{\boldsymbol{I}})$, where ${\sigma_n^2}$ is the noise power. At the receiver, ${\boldsymbol{\hat s}}$ is further fed into the JSCC decoder $f_d$ to reconstruct the latent representation $\boldsymbol{\hat y}$, which is further used by the nonlinear synthesis transform $g_s$ to recover the source data $\boldsymbol{\hat x}$. The whole procedure of NTSCC system is
\begin{equation}\label{eq_trans_process}
\begin{aligned}
& {{\boldsymbol{x}}} \xrightarrow{{{g_a}( \cdot ;{\boldsymbol{\phi}}_g)}} {{\boldsymbol{y}}}
\xrightarrow{{{f_e}( \cdot ;{\boldsymbol{\phi}}_f)}} {{\boldsymbol{s}}}
\xrightarrow{{{W}( \cdot|\boldsymbol{h} )}} {{\boldsymbol{\hat s}}}
\xrightarrow{{{f_d}( \cdot ;{\boldsymbol{\theta}}_f)}} {{\boldsymbol{\hat y}}}
\xrightarrow{{{g_s}( \cdot ;{\boldsymbol{\theta}}_g)}} {{\boldsymbol{\hat x}}} \\
& \text{with the latent prior~} {{\boldsymbol{y}}} \xrightarrow{{{h_a}( \cdot ;{\boldsymbol{\phi}}_h)}} {{\boldsymbol{z}}}
\xrightarrow{{{h_s}( \cdot ;{\boldsymbol{\theta}}_h)}} \left\{ {{\boldsymbol{\mu}},{\boldsymbol{\sigma}}} \right\},
\end{aligned}
\end{equation}
where $(\boldsymbol{\phi}, \boldsymbol{\theta}) = (\boldsymbol{\phi}_g,\boldsymbol{\phi}_h,\boldsymbol{\phi}_f,\boldsymbol{\theta}_g,\boldsymbol{\theta}_h,\boldsymbol{\theta}_{f})$ encapsulate learnable network parameters of each function. The system efficiency is measured by the \emph{channel bandwidth ratio (CBR)} $\rho = k/m$.

The key of NTSCC lies in the \emph{variable-rate} JSCC guided by the latent prior on the semantic feature space. The latent prior $p_{\boldsymbol{y} | \boldsymbol{z}} (\boldsymbol{y} | \boldsymbol{ z})$ is obtained as
\begin{equation}\label{eq_ntscc_z_to_y}
\begin{aligned}
  p_{\boldsymbol{y} | \boldsymbol{z}} (\boldsymbol{y} | \boldsymbol{ z}) = & \prod_i \underbrace{\left( \mathcal{N}(y_i|{{\mu}}_i,{{\sigma}}_i^2) * \mathcal{U}(-\frac{1}{2},\frac{1}{2}) \right)}_{p_{y_i | \boldsymbol{z}}} ({ y}_i) \\
  ~ & \text{with~} (\boldsymbol{ \mu},\boldsymbol{ \sigma}) = {h_s}(\boldsymbol{ z}; \boldsymbol{\theta}_h),
\end{aligned}
\end{equation}
where the convolutional operation ``$*$'' with a standard uniform distribution is used to match the prior to the marginal such that the estimated rate $-\log{p_{\boldsymbol{y} | \boldsymbol{z}} (\boldsymbol{y} | \boldsymbol{ z})}$ is non-negative. Different from the data compression where the uniformly-noised proxy ${\tilde y}_i$ is applied in \eqref{eq_ntscc_z_to_y}, we do not need the quantization operation such that the raw $y_i$ can be used. The hyperprior $\boldsymbol{z}$ is usually transmitted over the digital link as side information due to its small cost, where the quantization $\lfloor \cdot \rceil$ (rounding to integers) is needed. A uniformly-noised proxy $\boldsymbol{\tilde z} = \boldsymbol{z} + \boldsymbol{o}$ is used to replace the quantized representation $\boldsymbol{\bar y} = \lfloor \boldsymbol{y} \rceil$ during model training \cite{balle2018}, where $o_j$ is sampled from $\mathcal{U}(-\frac{1}{2},\frac{1}{2})$. The probability of hyperprior $\boldsymbol{\tilde z}$ is calculated on the fully factorized density $p_{\boldsymbol{z}} = \prod\nolimits_j p_{{z}_j}$ as
\begin{equation}\label{eq_entropy_model_z}
  p_{\boldsymbol{z}} (\boldsymbol{\tilde z}) = \prod_j  \underbrace{\left( p_{{z}_j | \boldsymbol{\psi}^{(j)}} ({z}_j | \boldsymbol{\psi}^{(j)}) * \mathcal{U}(-\frac{1}{2},\frac{1}{2}) \right)}_{p_{{z}_j}} ({\tilde z}_j),
\end{equation}
where $\boldsymbol{\psi}^{(j)}$ encapsulates all the parameters of $p_{{z}_j | \boldsymbol{\psi}^{(j)}}$.

The optimizing problem of NTSCC is formulated following the variational inference context \cite{balle2018}, the posterior distribution $p_{\boldsymbol{\hat s}, {\boldsymbol{\tilde z}} | \boldsymbol{x}}$ is approximated using the variational density $q_{\boldsymbol{\hat s}, {\boldsymbol{\tilde z}} | \boldsymbol{x}}$ by minimizing their Kullback-Leibler (KL) divergence over the data distribution $p_{\boldsymbol{x}}$ and the channel gain distribution $p_{\boldsymbol{h}}$ as the equation (11) in \cite{dai2022nonlinear}. Accordingly, the optimization of NTSCC system can be converted to the minimization of the expected channel bandwidth cost, as well as the expected distortion of the reconstructed data versus the original, which leads to the optimization of the following end-to-end RD tradeoff,
\begin{equation}\label{eq_expect_loss_func}
\begin{aligned}
  & \mathcal{L}_{\text{RD}}(\boldsymbol{\phi}, \boldsymbol{\theta}, \boldsymbol{\psi}) = \mathbb{E}_{\boldsymbol{x}\sim p_{\boldsymbol{x}}}\mathbb{E}_{\boldsymbol{h}\sim p_{\boldsymbol{h}}} D_{\rm{KL}}(q_{\boldsymbol{\hat s},\boldsymbol{\tilde z} | \boldsymbol{x}} \| p_{\boldsymbol{\hat s},\boldsymbol{\tilde z} | \boldsymbol{x}} ) \Leftrightarrow \mathbb{E}_{\boldsymbol{x}\sim p_{\boldsymbol{x}}}  \\
  & \mathbb{E}_{\boldsymbol{h}\sim p_{\boldsymbol{h}}} \Big( \underbrace{ -{\eta_y} \log{p_{\boldsymbol{ y}|\boldsymbol{ z}}(\boldsymbol{ y}|\boldsymbol{ z})} - {\eta_z}{\log{p_{\boldsymbol{z}}(\boldsymbol{\tilde z})}} }_{\text{channel bandwidth cost:~}R} + \lambda \cdot \underbrace{d(\boldsymbol{x},\boldsymbol{\hat{x}})}_{\text{distortion:~}D}\Big),
\end{aligned}
\end{equation}
where the \emph{Lagrange multiplier} $\lambda$ on the end-to-end distortion term determines the tradeoff between the transmission rate $R$ and the end-to-end distortion $D$. The \emph{bandwidth scaling factors} $\eta_y$ and $\eta_z$ are from the estimated entropy to the channel bandwidth cost and tied with the source-channel codec capability and the wireless channel state. A larger $\eta_y$ indicates a better performance on JSCC codec $f_e$ and $f_d$, but incurs more channel bandwidth cost. Accordingly, $\eta_y$ can also be adjusted as a hyperparameter to control the system RD tradeoff. $\eta_z$ cannot be adjusted manually since explicit entropy coding \cite{witten1987arithmetic} and LDPC coding \cite{richardson2018design} are selected to transmit the side information $\bm{z}$ reliably. Notably, different from that in the data compression settings which use the codebook-based arithmetic coding, the side information $\bm{z}$ in NTSCC is not necessary for the receiver as analyzed in \cite{dai2022nonlinear}. If $\bm{z}$ is not transmitted, the decoding performance shows some degradation while the bandwidth cost is also reduced. Overall, the end-to-end RD performance is comparable. Therefore, in the subsequent context, we do not transmit the side information $\bm{z}$ for simplicity, thereby the bandwidth cost term of size information in \eqref{eq_expect_loss_func} is omitted, and the only bandwidth scaling factor $\eta_y$ is abbreviated as $\eta$.

In practice, each embedding $y_i$ is a $C$-dimensional feature vector. The learned entropy model $-\log p_{{ y}_i|\boldsymbol{ z}}({ y}_i|\boldsymbol{ z})$ indicates the summation of entropy along $C$ channels of $y_i$, thus, the information density distribution of $\boldsymbol{y}$ is captured. Accordingly, the bandwidth cost, such as the number of OFDM subcarriers, ${\bar k}_{i}$ for transmitting $y_i$ can be determined as
\begin{equation}\label{eq_channel_bandwidth_cost_cal}
  {\bar k}_{i} = Q({k}_{i}) = Q\Big( {-{\eta} \log{p_{y_i|\boldsymbol{ z}}(y_i|\boldsymbol{ z})}} \Big),
\end{equation}
where the learned entropy model $p_{{y}_i|\boldsymbol{ z}}$ follows \eqref{eq_ntscc_z_to_y}, $Q$ denotes a scalar quantization whose range includes $2^{q}$ ($q = 1,2,\dots$) integers, and the quantization set is related to the bandwidth scaling factor $\eta$ and the Lagrange multiplier $\lambda$. Hence, the predetermined $q$ bits is transmitted as extra side information to inform the receiver which bandwidth is allocated to every embedding $y_i$. To adaptively map $y_i$ to a ${\bar k}_{i}$-dimensional channel-input vector $s_i$, the dynamic neural network structure is introduced into Transformers \cite{dosovitskiy2020image} to realize the variable-rate JSCC codec $f_e$ and $f_d$ \cite{dai2022nonlinear}. \emph{This variable-rate JSCC coding mechanism applied to each latent feature $y_i$ is indeed the main source of NTSCC performance gain versus the plain deep JSCC, which constructs an explicit match between the coded transmission resource allocation and the source content complexity.}

\section{Contextual NTSCC: More Efficient SCT}\label{section_context}

\begin{figure}[t]
	\setlength{\abovecaptionskip}{0.cm}
	\setlength{\belowcaptionskip}{-0.cm}
	\centering{\includegraphics[scale=0.5]{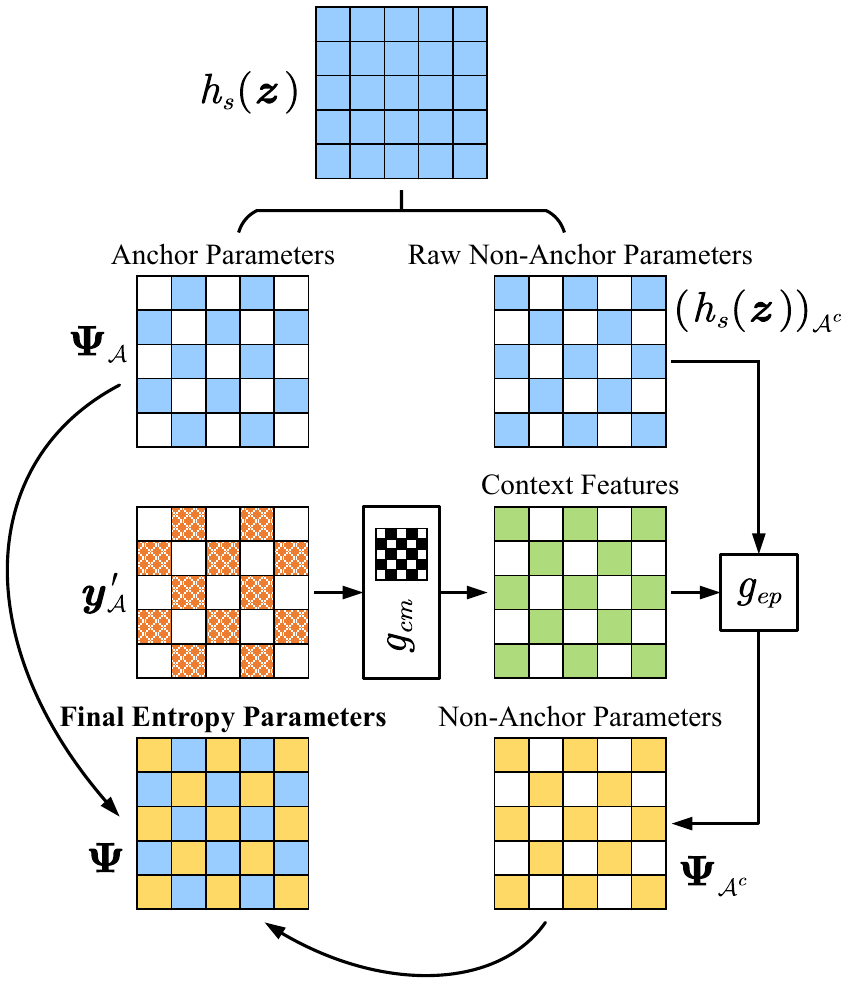}}
	\caption{Illustration of the checkerboard context model using the $5 \times 5$ convolutional kernel masked by the checkerboard-shaped binary mask. All colored blocks denote the effective positions.}\label{Fig_checkerboard}
	\vspace{0em}
\end{figure}

As analyzed in Section \ref{section_introduction}, there are still many problems
to solve before NTSCC models can be deployed in practice. From this section, we propose a comprehensive framework including three improvements to derive the \emph{improved NTSCC}. The whole improved NTSCC system is illustrated in Fig. \ref{Fig_ntsccpp_system_architecture}.

From the above descriptions of NTSCC, we can find that the RD performance is determined partially by the entropy model, a more accurate distribution estimation of latent code $\bm{y}$ will definitely lead to more accurate bandwidth allocation $k_i$ for each latent feature $y_i$, thereby a more efficient variable-rate JSCC can be formed. In the initial work of NTSCC \cite{dai2022nonlinear}, the correlations among latents $\{ y_i \}$ is modeled to some extent using the \emph{hierarchical prior}, i.e., the parameterized entropy model of $\bm{y}$ is built with the hyperprior $\bm{z} = h_a(\bm{y})$ as \eqref{eq_ntscc_z_to_y}. Recent frameworks based on the \emph{context model} are emerging to capture the correlations among the latents $\{ y_i \}$ more efficiently \cite{minnen2018,lee2018context,cheng2020learned,choi2019variable}. It intuitively provides an idea to further boost the end-to-end RD performance in NTSCC.

Accordingly, to estimate the location and scale parameters $\bm{\Psi} = (\bm{\mu},\bm{\sigma})$ of the entropy model for latent $\bm{y}$ in \eqref{eq_ntscc_z_to_y}, we can further introduce a parameter inference function $g_{ep}$. Let $h_s(\bm{z})$ denote the hyperprior feature and $g_{cm}(\bm{y}_{<i})$ denote the context feature, the entropy model parameter prediction for $i$-th latent code $y_i$ is formulated as
\begin{equation}\label{eq_autoregessive_model}
  {\bm{\Psi}}_i = (\mu_i,\sigma_i) = g_{ep}(h_s({\bm{z}}),g_{cm}(\bm{y}_{<i})),
\end{equation}
where $\bm{y}_{<i}$ means the \emph{causal context}, i.e., some visible decoded latents of the latent $y_i$. Let $\bm{\omega} = (\bm{\omega}_{cm}, \bm{\omega}_{ep})$ encapsulate the learnable network parameters of functions $g_{cm}$ and $g_{ep}$. Under this context setting, the optimization goal of NTSCC system is reformulated as
\begin{equation}\label{eq_expect_context_loss_func}
\begin{aligned}
  & \mathcal{L}_{\text{RD}}(\boldsymbol{\phi}, \boldsymbol{\theta}, \boldsymbol{\psi}, \boldsymbol{\omega}) = \\
  & \mathbb{E}_{\boldsymbol{x}\sim p_{\boldsymbol{x}}} \mathbb{E}_{\boldsymbol{h}\sim p_{\boldsymbol{h}}} \Big( -{\eta} \log{p_{\boldsymbol{ y}|\boldsymbol{ z}}(\boldsymbol{ y}|\boldsymbol{ z})} + \lambda \cdot {d(\boldsymbol{x},\boldsymbol{\hat{x}})}\Big),
\end{aligned}
\end{equation}
the context entropy model $p_{\boldsymbol{y} | \boldsymbol{z}} (\boldsymbol{y} | \boldsymbol{ z})$ is built as
\begin{equation}\label{eq_ntscc_z_to_y_context}
\begin{aligned}
  p_{\boldsymbol{y} | \boldsymbol{z}} (\boldsymbol{y} | \boldsymbol{ z}) = \prod_i \underbrace{\left( \mathcal{N}(y_i|{{\mu}}_i,{{\sigma}}_i^2) * \mathcal{U}(-\frac{1}{2},\frac{1}{2}) \right)}_{p_{y_i|\bm{z}}} ({ y}_i)
\end{aligned}
\end{equation}
where each parameter tuple ${\bm{\Psi}}_i = (\mu_i,\sigma_i)$ is obtained from \eqref{eq_autoregessive_model}. In this way, the context and hierarchical
priors are combined to estimate the distribution of $\bm{y}$.

An intuitive implementation of $g_{cm}(\bm{y}_{<i})$ is the autoregressive manner \cite{minnen2018}, where $\bm{y}_{<i}$ denotes the set of all latents with indices smaller than $i$, i.e., $\bm{y}_{<i} = (y_1,y_2,\dots,y_{i-1})$. Despite its good performance, this \emph{serial} context model has a severe limitation on its computational efficiency which cannot be employed in practice. Therefore, in this paper, we follow the idea of \cite{he2021checkerboard} to build the context model in a \emph{checkerboard} manner, it is helpful to building a computationally efficient parallel context model without introducing apparent performance loss compared with the serial context model.

Following the checkerboard shaped context \cite{he2021checkerboard}, we only encode/decode half of the latents, using checkerboard-shaped context and hyperprior. The coding of the other half of latents, which we call \emph{anchors}, only depends on the hyperprior. To implement the two types of context rules, the context feature of all anchors are set to zero, and the calculation of entropy parameters $\bm{\Psi}_i$ in \eqref{eq_autoregessive_model} to a spatial location conditioned form:
\begin{equation}\label{eq_checkerboard_model}
  \bm{\Psi}_i = \left\{ \begin{aligned}
& h_s(\bm{z}), \text{~for~} i \in {{\mathcal{A}}},\\
& g_{ep}(h_s({\bm{z}}),g_{cm}(\bm{y}_{\mathcal{A}}^\prime;\bm{M} \odot \bm{W}))_i, \text{~for~} i \in {\mathcal{A}^{c}},
\end{aligned} \right.
\end{equation}
where $\mathcal{A}$ denotes the set of anchor indices, its complementary set $\mathcal{A}^c$ denotes the non-anchor indices, thereby ${\bm{y}}_{\mathcal{A}}$ denotes the set of latent codes whose indices belong to $\mathcal{A}$. A special note here, since the transmitter does not know the reconstructed $\bm{\hat{y}}_{\mathcal{A}}$ by JSCC decoder at the receiver, $\bm{{y}}_{\mathcal{A}}^\prime$ in \eqref{eq_checkerboard_model} is an estimate of $\bm{\hat{y}}_{\mathcal{A}}$ based on the statistics of the channel. We assume the transmitter has a local copy of the decoder parameters. In order to
generate $\bm{{y}}_{\mathcal{A}}^\prime$, the transmitter simulates locally independent realizations of the channel and the decoder models, obtaining
\begin{equation}\label{eq_simulated_y}
  \bm{{y}}_{\mathcal{A}}^\prime = \frac{1}{n}\sum_{i=1}^n \bm{\tilde{y}}_{\mathcal{A}}^{(i)},
\end{equation}
where $\bm{\tilde{y}}_{\mathcal{A}}^{(i)}$ denotes is the $i$-th realization of the simulation of the transmitter's semantic latent feature reconstruction given $\bm{y}_{\mathcal{A}}$ as the input of variable-rate JSCC codec in the NTSCC pipeline in \eqref{eq_trans_process}, and $n$ is the total number of independent channel realizations used to estimate the receiver's output. In addition, $g_{cm}$ in \eqref{eq_checkerboard_model} denotes the masked convolution operation conditioned on a checkerboard-shaped mask, defined as
\begin{equation}
  g_{cm}(\bm{x}) = (\bm{M} \odot \bm{W})\bm{x} + \bm{b},
\end{equation}
where $\bm{W}$ denotes the $l \times l$ convolutional weights, $\bm{M}$ is the $l \times l$ binary mask describing the context modeling pattern, $\bm{b}$ is a learned bias term. In this paper, we employ the $5 \times 5$ convolutional kernel $\bm{W}$ masked by the checkerboard-shaped binary mask $\bm{M}$. Under this setting, the resulted entropy model parameters belonging to the anchor branch $\bm{\Psi}_{\mathcal{A}}$ and the non-anchor branch $\bm{\Psi}_{\mathcal{A}^c}$ are obtained as shown in Fig. \ref{Fig_checkerboard}.

During encoding, all latents in $\bm{y}$ are visible, we thus can calculate all entropy model parameters $\bm{\Psi} = (\bm{\Psi}_{\mathcal{A}}, \bm{\Psi}_{\mathcal{A}^c})$ with only one pass of context model. Finally, the variable-rate JSCC encodes anchor latents $\bm{y}_{\mathcal{A}}$ and non-anchor latents $\bm{y}_{\mathcal{A}^c}$ into the channel-input symbol stream $\bm{s}$ as that in the original NTSCC \cite{dai2022nonlinear,wang2023wireless}. The $i$-th component ${{s}}_i$ of $\bm{s}$ is a symbol vector whose length is determined as ${\bar{k}}_i$ in \eqref{eq_channel_bandwidth_cost_cal},
where the context entropy model $p_{{ y}_i|\boldsymbol{ z}}$ follows \eqref{eq_ntscc_z_to_y_context}. During the contextual JSCC encoding, its pipeline includes two streams, i.e., anchor and non-anchor parts, as demonstrated in Fig. \ref{Fig_ntsccpp_system_architecture} (here some notations will be discussed in the next Section \ref{section_versatile} which serves for building the bandwidth compatible codec), their corresponding sub-encoders $f_{e_a}$ and $f_{e_{na}}$ are parameterized by $\bm{\phi}_{f_{a}}$ and $\bm{\phi}_{f_{na}}$, respectively. The estimated anchor context latent codes $\bm{y}_{\mathcal{A}}^\prime$ are employed as the conditional input to the JSCC encoding of the non-anchor branch $\bm{y}_{\mathcal{A}^c}$. These two encoding pipelines are carried out in \emph{one pass}.

In the receiver, two decoding passes are needed to obtain the whole $\bm{\hat y}$ as depicted in Fig. \ref{Fig_ntsccpp_system_architecture}. The JSCC sub-decoder $f_{d_{a}}$ (parameterized by $\bm{\theta}_{f_a}$) will be first executed for the anchor branch to obtain $\bm{\hat{y}}_{\mathcal{A}}$. Then the decoded anchor latent codes $\bm{\hat{y}}_{\mathcal{A}}$ are visible in the next decoding pass, which are fed as the conditional input to aid the decoding of the non-anchor branch in $f_{d_{na}}$ (parameterized by $\bm{\theta}_{f_{na}}$). Overall, the JSCC decoding is a \emph{two pass} procedure, we thus build the new variable-rate \emph{contextual JSCC codec} aligned with the checkerboard context entropy model. The whole \emph{contextual NTSCC} system is dubbed ``NTSCC+'' in this paper.

\section{Compatible NTSCC: Single Model, Wide Use}\label{section_versatile}

In this section, we consider a curial issue related to the practical deployment of NTSCC based end-to-end communication system. No matter for the standard or the contextual NTSCC models, the end-to-end RD tradeoff is typically parameterized by the hyperparameter $\lambda$ on the distortion term as \eqref{eq_expect_loss_func} and \eqref{eq_expect_context_loss_func} (named as distortion weight). The distortion weight $\lambda$ plays an important role in training NTSCC models and requires careful tuning for various applications. Moreover, the learning procedure of NTSCC is tied with channel conditions, which enables the model to be compatible for channel fading, noise, etc. By training multiple models with different values of distortion weight $\lambda$ and channel SNR $\nu$, we can obtain different points on a rate-SNR-distortion surface. Unfortunately, as rate-SNR-distortion surfaces depend on the dataset, channel conditions, and architecture, practitioners need to tune $\lambda$ and $\nu$ for each individual semantic communication task, which will bring large training cost and model storage burden. Such shortcoming calls for a mechanism supporting neural transformation and JSCC codec to achieve wide rate range and channel SNRs in a single model. In this work, we introduce an improved NTSCC framework that does not require hyperparameters tuning on $\lambda$ and $\nu$, and can learn to support different rates and channel SNRs in a single training run. Our approach is named \emph{compatible NTSCC}.

\subsection{Key Insights to Motivate Our Divide-and-Conquer Design}

Our problem is inherently equivalent to formulate \emph{response functions} \cite{ha2016hypernetworks} $\bm{\phi}^{*}(\lambda, \nu)$ and $\bm{\theta}^{*}(\lambda, \nu)$, which map the distortion weight $\lambda$ and the channel SNR $\nu$ to the optimal transformation and JSCC codec parameters trained with such $\lambda$ and $\nu$. Next, through some key experimental observations, we justify that the response function $\bm{\phi}^{*}(\lambda, \nu)$ in NTSCC can be decoupled as the $\bm{\phi}_g^{*}(\lambda)$ operated on the analysis transform and the $\bm{\phi}_f^{*}(\nu)$ operated on the JSCC encoder. Similarly, $\bm{\theta}^{*}(\lambda, \nu)$ can be decoupled as $\bm{\theta}_g^{*}(\lambda)$, and $\bm{\theta}_f^{*}(\nu)$.

\begin{figure}[t]
\setlength{\abovecaptionskip}{0.cm}
\setlength{\belowcaptionskip}{-0.cm}
 \centering
 \subfigure[]{\includegraphics[width=0.24\textwidth]{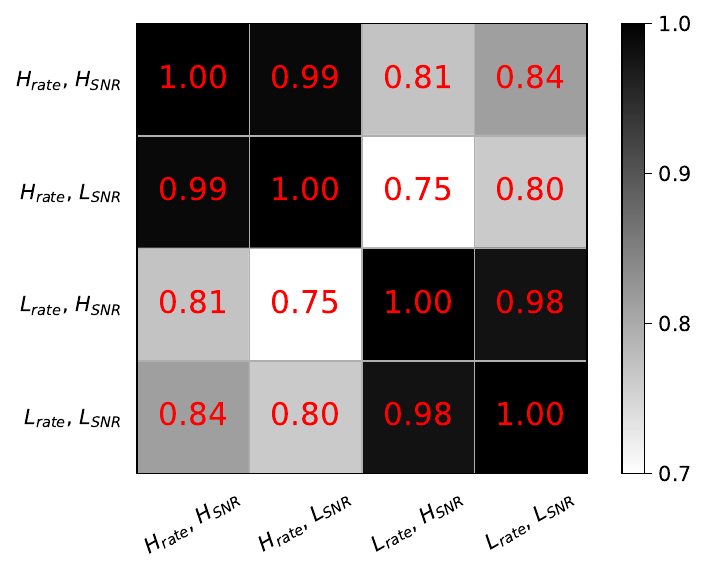}}
 \subfigure[]{\includegraphics[width=0.24\textwidth]{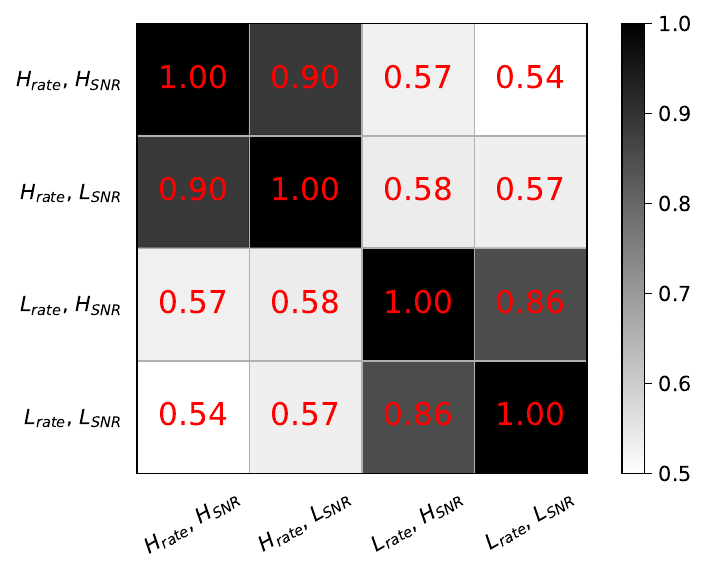}}
 \caption{Averaged cosine similarity matrix on Kodak image dataset for (a) $\bm{\dot{y}}$ and (b) $\bm{\dot{v}}$. In implementation, we finetuned the ($H_{\text{rate}}, H_{\text{SNR}}$) model to the other three models by varying the distortion weight $\lambda$ or the channel SNR $\nu$.}
 \label{Fig_insights}
 \vspace{0em}
\end{figure}

As marked in Fig. \ref{Fig_ntsccpp_system_architecture}, with some notation abuse, we monitor the change rule of $\bm{\dot{y}} = g_a(\bm{x})$ (latent features output from the semantic analysis transform) and $\bm{\dot{v}} = f_e(\bm{y})$ (JSCC codewords before rate-matched to the channel-input sequence $\bm{s}$) when we change the model learning hyperparameters $\lambda$ and $\nu$. The results are provided in Fig. 3, where we calculate the average of cosine similarity among 4 individually trained models using the combination of two different distortion weights and two different channel SNRs. $H_{\text{rate}}$ and $L_{\text{rate}}$ denote NTSCC models trained under $\lambda = 0.72$ and $\lambda = 0.18$, respectively. $H_{\text{SNR}}$ and $L_{\text{SNR}}$ mark NTSCC models trained under $\nu = 10$dB and $\nu = 0$dB, respectively. For brevity, we use the combination of two hyperparameters to represent each model. Specifically, Fig. 3(a) shows the cosine similarity between latent feature maps $\bm{\dot{y}}$, Fig. 3(b) shows the cosine similarity between JSCC unrate-matched codewords $\bm{\dot{v}}$. From the results, we can draw the following insights.
\begin{enumerate}
  \item Overall, when varying the model training hyperparameters $\lambda$ and $\nu$, the feature map $\bm{\dot{y}}$ and the JSCC codeword $\bm{\dot{v}}$ are both clearly correlated (cosine similarity $> 0.54$).

  \item According to Fig. 3(a), NTSCC models learned under the same rate (i.e., the CBR $\rho$) but the different channel SNR produce the \emph{extremely strong correlated} latent features $\bm{\dot{y}}$ (cosine similarity $0.99$ or $0.98$). That means the latent codes output by the analysis transform $g_a$ are almost invariant when we change the transmission channel condition. In other words, \emph{the parameters of $g_a$ can be shared among models applied to different channel SNRs with negligible performance loss.}

  \item According to Fig. 3(b), NTSCC models learned under the same rate (i.e., the CBR $\rho$) but the different channel SNR produce the \emph{reasonably high correlated} JSCC codewords $\bm{\dot{v}}$ (cosine similarity $0.90$ or $0.86$). That means \emph{the outputs of JSCC encoder $f_e$ under different SNRs could be transformed among each other.}

  \item Jointly considering the results in Fig. 3(a) and 3(b), we can justify that \emph{it will be much more efficient to implement the rate adjustment by introducing the transformation on the latent codes $\bm{\dot{y}}$ than on the JSCC codewords $\bm{\dot{v}}$.} When comparing the models learned under different rates but the same channel SNR in Fig. 3(a) and 3(b), we find that the correlation between the JSCC codewords $\bm{\dot{v}}$ (cosine similarity $0.57$) is apparently weaker than that of the latent features $\bm{\dot{y}}$ (cosine similarity $0.80$ or $0.81$). This observation supports the above insight.
\end{enumerate}

In a comprehensive manner, we can justify that the response function $\bm{\phi}^{*}(\lambda, \nu)$ in NTSCC can be decoupled as $\bm{\phi}_g^{*}(\lambda)$ and $\bm{\phi}_f^{*}(\nu)$, and the similar way for decoupling $\bm{\theta}^{*}(\lambda, \nu)$. This interesting insight aligns well with the design philosophy of our traditional separated source and channel codecs, where the linear transform coding (LTC) of source codec varies the quantization step size for the rate adjustment, and the channel coded modulation adapts to different channel SNRs via the adaptive modulation coding (AMC) mechanism. Similarly, in our compatible NTSCC based SCT system, the rate adaptation is also carried out by the nonlinear transformation modules while the channel adaptation is performed on the JSCC codec. As marked in Fig. \ref{Fig_ntsccpp_system_architecture}, we can employ the distortion weight $\lambda$ as the source quality indicator (SQI) in the transmitter to adjust the average channel bandwidth cost, and the average SNR $\nu$ as the channel quality indicator (CQI) to match the average channel SNR estimated at the receiver. Both SQI and CQI are 16bit floating-point numbers which are shared in the transceiver via side information link
and feedback link, respectively. Next, we discuss how to leverage SQI and CQI to build the response functions $\bm{\phi}_g^{*}(\lambda)$, $\bm{\phi}_f^{*}(\nu)$, $\bm{\theta}_g^{*}(\lambda)$, and $\bm{\theta}_f^{*}(\nu)$.

\subsection{Response Networks for Compatible NTSCC+}

In this part, based on the contextual NTSCC (NTSCC+) framework proposed in Section \ref{section_context}, we design the approaches for explicitly modeling the response functions with the dual-granularity code scaling mechanism. The whole architecture is abbreviated as ``compatible NTSCC+''.

\subsubsection{Rate Adjustment in Compatible NTSCC+}

As analyzed before, to adjust the overall rate, the response function $\bm{\phi}_g^{*}(\lambda)$ can be built by directly manipulating the latent feature $\bm{\dot{y}}$ output from the well-trained semantic analysis transform $g_a$ which is parameterized with $\bm{\phi}_{g,\lambda_0,\nu_0}$. Herein, $\bm{\phi}_{g,\lambda_0,\nu_0}$ encapsulates the network parameters learned under the default hyperparameters $\lambda_0$ and $\nu_0$. In general, due to the highly-correlated latent codes $\bm{\dot{y}}$ given by models under different rates, we can apply a well-learned affine transform to transfer from one baseline latent code to the other target latent codes \cite{balle2020nonlinear}. This affine transform is controlled by the hyperparameter $\lambda$. For brevity, we construct a compact approximation of the affine transform to build the response function $\bm{\phi}_g^{*}(\lambda)$ as $\bm{\phi}_g^{*}(\lambda) = (\bm{\phi}_{g,\lambda_0,\nu_0}, \bm{q}_{\rm{rate}})$,
where the \emph{channel-wise scaling vector} $\bm{q}_{\rm{rate}}^{(i)}$ ($C$ dimensions) applied to the $C$-dimensional latent vector ${\dot{y}}_i$ output from the $\bm{\phi}_{g,\lambda_0,\nu_0}$ parameterized $g_a$ is written as
\begin{equation}\label{eq_q_g}
  \bm{q}_{\rm{rate}}^{(i)} = \left\{ \begin{aligned}
& \bm{q}_{a,\rm{rate}} = r_{\rm{global},\lambda} \cdot \bm{r}_{a},\text{~for~} i\in \mathcal{A},\\
& \bm{q}_{na,\rm{rate}} = r_{\rm{global},\lambda} \cdot \bm{r}_{na},\text{~for~} i\in \mathcal{A}^c,
\end{aligned} \right.
\end{equation}
which is aligned with the operations in \cite{choi2019variable} and \cite{li2022hybrid}. From \eqref{eq_q_g}, we find that our dual-granularity approach involves two different kinds of scaling operations on both the anchor and the non-anchor branches: the global scaling scalar $r_{\rm{global},\lambda}$, and the channel-wise scaling vectors $\bm{r}_{a}$ and $\bm{r}_{na}$. The $r_{\rm{global},\lambda}$ is only a single value and is set from the SQI $\lambda$ for controlling the overall target rate. As all elements of $\bm{\dot{y}}$ are manipulated by the same $r_{\rm{global},\lambda}$, it brings a coarse rate adjustment effect. Motivated by the channel attention mechanism \cite{hu2018squeeze}, $\lambda$-independent additional scaling vectors $\bm{r}_{a}$ and $\bm{r}_{na}$ are learned to scale $\bm{\dot{y}}$ at different channels to match finely with their channel-wise semantic importance differences. Under this setting, we build a new \emph{response network} for the semantic analysis transform $g_a^{*}$ that is parameterized by the response function $\bm{\phi}_g^{*}(\lambda)$. As marked in Fig. \ref{Fig_ntsccpp_system_architecture}, $g_a^{*}(\bm{x})$ is formulated as
\begin{equation}\label{eq_new_ga}
    g_a^{*}(\bm{x}) = \bm{y} = \left\{ {\dot{y}}_i \oslash \bm{q}_{\rm{rate}}^{(i)} \right\} = \left\{ \begin{aligned}
& {\dot{y}}_i \oslash \bm{q}_{a,\rm{rate}},\text{~for~} i\in \mathcal{A},\\
& {\dot{y}}_i \oslash \bm{q}_{na,\rm{rate}},\text{~for~} i\in \mathcal{A}^c,
\end{aligned} \right.
\end{equation}
herein, $i$ marks the spatial position, $\oslash$ denotes the element-wise division, and with some notation abuse, ${\dot{y}}_i$ represents the $C$-dimensional vector along the channels at the $i$-th spatial position of the tensor $\bm{\dot{y}}$. In such a way, the final latents output from the new response semantic analysis transform $\bm{y} = g_a^{*}(\bm{x})$ is controlled by the SQI $\lambda$.

Correspondingly, as depicted in Fig. \ref{Fig_ntsccpp_system_architecture}, in the receiver, we also build the response semantic synthesis transform $g_s^{*}(\bm{\hat{y}})$ using the response function $\bm{\theta}_g^{*}(\lambda) = (\bm{\theta}_{g,\lambda_0,\nu_0}, \bm{q}_{\rm{rate}})$, we thus write the reconstructed version of $\bm{{\dot{y}}}$ as
\begin{equation}
  \bm{\hat{\dot{y}}} = \left\{ \hat{{y}}_i \odot \bm{q}_{\rm{rate}}^{(i)} \right\} =  \left\{ \begin{aligned}
& \hat{{y}}_i  \odot \bm{q}_{a,\rm{rate}},\text{~for~} i\in \mathcal{A},\\
& \hat{{y}}_i  \odot \bm{q}_{na,\rm{rate}},\text{~for~} i\in \mathcal{A}^c.
\end{aligned} \right.
\end{equation}
The final reconstructed signal is $\bm{\hat{x}} = g_s^{*}(\bm{\hat{y}}) = g_s(\bm{\hat{\dot{y}}})$, where $g_s$ is parameterized by $\bm{\theta}_{g,\lambda_0,\nu_0}$.

\begin{figure}[t]
\setlength{\abovecaptionskip}{0.cm}
\setlength{\belowcaptionskip}{-0.cm}
 \centering
 \subfigure[]{\includegraphics[height=95pt]{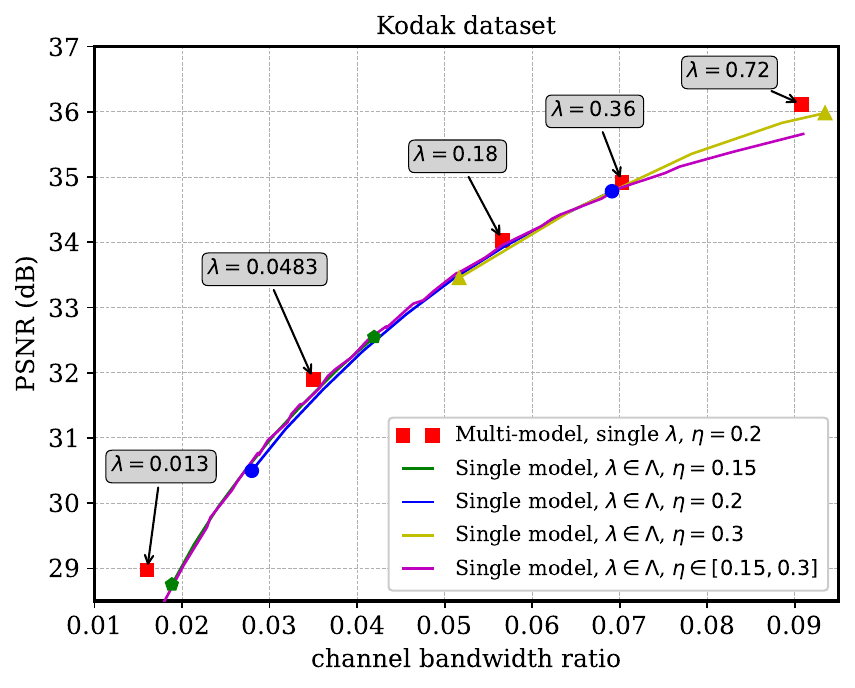}}
 \subfigure[]{\includegraphics[height=95pt]{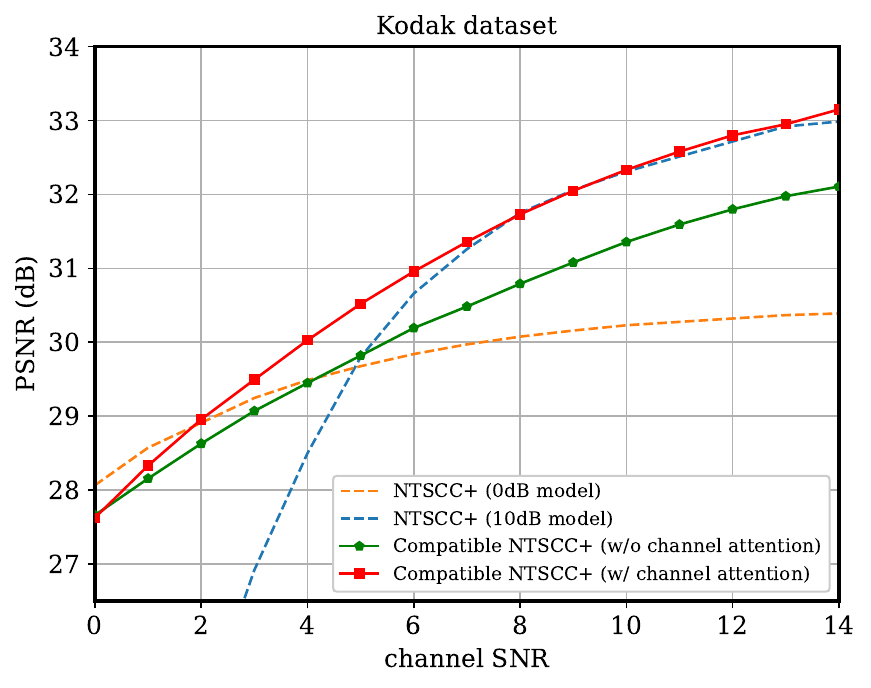}}
 \caption{Useful insights to guide the design of response networks. (a) Rate-distortion (RD) curve of compatible NTSCC+ over the additive white Gaussian noise (AWGN) channel at SNR $=10$dB. Specifically,
  the ``multi-model'' scheme consists of 5 contextual NTSCC models, which are trained individually using the labelled $\lambda$ values. The ``single-model'' schemes are trained using the discrete $\lambda$ within the set $\Lambda=\{0.013, 0.0483, 0.18, 0.36, 0.72\}$. Besides, the bandwidth scaling factor $\eta$ used in training is provided in the legend, where the revised line denotes the model trained by using mixed $\eta$ values uniformly sampled from 0.15 to 0.3. (b) SNR-distortion curve of compatible NTSCC+ over the AWGN channel under the CBR constraint $\rho=0.04$.
 }
 \label{Fig_versatile_ntscc}
 \vspace{0em}
\end{figure}

We further investigate the capability of our new response nonlinear transform functions $g_a^{*}$ and $g_s^{*}$ by checking their adaptation range of the SQI $\lambda$ in a single model. As shown in Fig. \ref{Fig_versatile_ntscc}(a), each single model with fixed $\eta$ can very closely approach the multi-model RD curve with only slight performance loss, which verifies the effectiveness of our response networks. But we also find that their CBR covering range with one single model is somewhat limited. To further extend the compatible NTSCC coverage to the whole RD curve, we need to adjust the bandwidth scaling factor $\eta$ on each estimated entropy term $-\log p_{y_i|\bm{z}}(y_i)$. Compared to the SQI $\lambda$ for the fine channel bandwidth cost controlling, the bandwidth scaling factor $\eta$ is leveraged to coarsely manipulating the bandwidth cost range. Based on this insight, we suggest training the response network with cooperative use of $\lambda$ and $\eta$ values. As plotted with the revised line in Fig. \ref{Fig_versatile_ntscc}, it covers a broad range of the RD curve and aligns well with the multi-model scheme except for the very high rate region, which is due to the insufficient capacity in a single model.

\subsubsection{SNR Adaptation for Compatible NTSCC+}

Similar to the approach in the rate adjustment, we also construct the response function $\bm{\phi}_f^{*}(\nu) = (\bm{\phi}_{f_a,\lambda_0,\nu_0},\bm{\phi}_{f_{na},\lambda_0,\nu_0}, \bm{q}_{\rm{SNR}})$ to adapt a well-trained NTSCC+ model to various channel states, where the \emph{channel-wise scaling vector} $\bm{q}_{\rm{SNR}}^{(i)}$ applied to the JSCC codeword vector ${\dot{v}}_i$ is written as
\begin{equation}\label{eq_q_f}
  \bm{q}_{\rm{SNR}}^{(i)} = \left\{ \begin{aligned}
& \bm{q}_{a,\rm{SNR}} = c_{\rm{global},\nu} \cdot \bm{c}_{a},\text{~for~} i\in \mathcal{A},\\
& \bm{q}_{na,\rm{SNR}} = c_{\rm{global},\nu} \cdot \bm{c}_{na},\text{~for~} i\in \mathcal{A}^c.
\end{aligned} \right.
\end{equation}
where the $c_{\text{global},\nu}$ is the global scaling scalar associated with the CQI (usually adopted as the average channel SNR $\nu$), $\bm{c}_a$ and $\bm{c}_{na}$ are the channel-wise scaling vectors corresponding to the anchor and non-anchor positions, respectively. A special note, the CQI $\nu$ is preferred to be embedded as the channel attention vectors $\bm{q}_{a,\rm{SNR}}^{(i)}$ and $\bm{q}_{na,\rm{SNR}}^{(i)}$ related to the JSCC code vector $\dot{{v}}_i$, which enhances the capacity of our new response networks for the contextual JSCC codec. Accordingly, the modulation from $\dot{{v}}_i$ to $\bm{c}_{a}$ and $\bm{c}_{na}$ is a fully-connected network (FCN) including two linear layers and a rectied linear unit (ReLU) for non-linearity. The mapping procedures from $\nu$ and $\dot{\bm{v}}$ to $\bm{q}_{\rm{SNR}}$ are written as
\begin{equation}
\bm{q}_{\rm{SNR}}^{(i)} = \left\{ \begin{aligned}
& \bm{q}_{a,\rm{SNR}}^{(i)} = c_{\rm{global},\nu} \cdot \text{FCN}_a(\dot{{v}}_i),\text{~for~} i\in \mathcal{A},\\
& \bm{q}_{na,\rm{SNR}}^{(i)} = c_{\rm{global},\nu} \cdot \text{FCN}_{na}(\dot{{v}}_i),\text{~for~} i\in \mathcal{A}^c.
\end{aligned} \right.
\end{equation}

Under this setting, we can build the new \emph{response network} for the JSCC encoder $f_e^{*}$ (including $f_{e_a}^{*}$ and $f_{e_{na}}^{*}$) that is parameterized by the response function $\bm{\phi}_f^{*}(\nu)$. As marked in Fig. \ref{Fig_ntsccpp_system_architecture}, the unrate-matched JSCC codeword $\bm{v}$ given by the encoder $f_e^{*}(\bm{y})$ is formulated as
\begin{equation}\label{eq_new_ga}
  \bm{v} = \left\{ {\dot{v}}_i \oslash \bm{q}_{\rm{rate}}^{(i)} \right\} = \left\{ \begin{aligned}
& {\dot{v}}_i \oslash \bm{q}_{a,\rm{SNR}}^{(i)},\text{~for~} i\in \mathcal{A},\\
& {\dot{v}}_i \oslash \bm{q}_{na,\rm{SNR}}^{(i)},\text{~for~} i\in \mathcal{A}^c,
\end{aligned} \right.
\end{equation}
where $\bm{\dot{v}}_{\mathcal{A}} = f_{e_a}(\bm{y}_{\mathcal{A}})$ and $\bm{\dot{v}}_{\mathcal{A}^c} = f_{e_{na}}(\bm{y}_{\mathcal{A}^c},\bm{y}_{\mathcal{A}}^\prime)$, $i$ denotes the spatial position. Finally, according to the estimate entropy term $-\log{p_{y_i|\boldsymbol{ z}}(y_i|\boldsymbol{ z})}$ and the bandwidth scaling factor $\eta$, each codeword vector $v_i$ is rate-matched to the ${\bar{k}}_i$-dimensional channel-input symbol vector $s_i$ as \eqref{eq_channel_bandwidth_cost_cal}. In such way, the final variable-rate JSCC codeword output from the new response JSCC encoder $\bm{s} = f_e^{*}(\bm{y})$ is adjusted by the CQI $\nu$.

Correspondingly, as depicted in Fig. \ref{Fig_ntsccpp_system_architecture}, in the receiver, we also build the response JSCC decoder $f_d^{*}(\bm{\hat{s}})$ using the response function $\bm{\theta}_f^{*}(\nu) = (\bm{\theta}_{f_a,\lambda_0,\nu_0},\bm{\theta}_{f_{na},\lambda_0,\nu_0}, \bm{q}_{\rm{SNR}})$, we thus write the reconstructed version of $\bm{{\dot{v}}}$ as
\begin{equation}
  \bm{\hat{\dot{v}}} = \left\{ \hat{{v}}_i \odot \bm{q}_{\rm{SNR}}^{(i)} \right\} =  \left\{ \begin{aligned}
& \hat{{v}}_i  \odot \bm{q}_{a,\rm{SNR}}^{(i)},\text{~for~} i\in \mathcal{A},\\
& \hat{{v}}_i  \odot \bm{q}_{na,\rm{SNR}}^{(i)},\text{~for~} i\in \mathcal{A}^c.
\end{aligned} \right.
\end{equation}
Hence, the reconstructed semantic latent features are $\bm{\hat{y}}_{\mathcal{A}} = f_{d_a}(\bm{\hat{\dot{y}}}_{\mathcal{A}})$ and $\bm{\hat{y}}_{\mathcal{A}} = f_{d_{na}}(\bm{\hat{\dot{y}}}_{\mathcal{A}^c}, \bm{\hat{y}}_{\mathcal{A}})$.

In Fig. \ref{Fig_versatile_ntscc}(b), we justify the effectiveness of our channel attention method to generate the scaling vectors $\bm{q}_{a,\rm{SNR}}$ and $\bm{q}_{na,\rm{SNR}}$. If we simplify the channel attention mechanism as that adopted in the rate adjustment, $\bm{c}_{a}$ and $\bm{c}_{na}$ are degraded to be learned and shared spatially along the checkerboard (without (w/o) channel attention) which are independent with $\dot{\bm{v}}$. Our channel attention method enables $\bm{c}_{a}$ and $\bm{c}_{na}$ tied with the JSCC codeword instance $\dot{\bm{v}}$. Apparently, our codeword-dependent attention method (marked as w/ channel attention) compensates for the performance loss in Fig. \ref{Fig_versatile_ntscc}(b).

\section{Online Editable NTSCC: More Flexible SCT}\label{section_overfit}

Based on the above two improvements, we have built the NTSCC+ and \emph{compatible NTSCC+} system. Although it can be very successful in minimizing the given end-to-end \emph{expected RD} cost over a full source dataset and ergodic channel responses, existing NTSCC models yet only pay attention to optimizing an average low RD cost on the training set and the predetermined distortion metric. For a given data/channel sample, or when we introduce new semantic guidance, e.g., the human perceptual metric \cite{zhang2018unreasonable}, the region-of-interest (ROI) \cite{cai2019end} guidance in XR, etc., whose corresponding RD loss function has not been seen during the model training, such a learned semantic codec may not be good at capturing the specific data semantic feature and channel state at this instance, resulting in suboptimal transform and coding during the model inference stage.

To tackle this, we further introduce a class of enhancement methods to the NTSCC+ system by \emph{online optimizing the semantic latent code and the variable-length JSCC codec individually}, on a per data sample and channel response basis, during the model inference stage. In other words, we turn to minimize the end-to-end \emph{instant RD} cost for substantial gains
on every data and target instance. This \emph{instance-adaptive} system is dubbed ``NTSCC++''. The key idea leverages the \emph{overfitting property} of neural networks \cite{lu2020content,yang2020improving,guo2020variable,liu2021overfitting,gao2022flexible,dai2022adaptive}. With the online semantic latent code editing and JSCC codec updating, we can empower SCT with more flexible coded transmission capabilities. In addition, new semantic guidance can extend the optimization target for NTSCC+, we can thus achieve different coded transmission requirements. Motivated by \cite{gao2022flexible}, as two examples, we explore the potential of online adaptation in ROI-based NTSCC++ and
multi-distortion optimized NTSCC++.

\subsection{Online Adaptation for Standard NTSCC++}


Based on the amortized model $(\boldsymbol{\phi}, \boldsymbol{\theta}, \boldsymbol{\psi}, \boldsymbol{\omega})$ learned on the entire dataset and ergodic channel states, given any specific source sample $\bm{x}$, we further optimize the goal \eqref{eq_instant_context_loss_func} by online updating the semantic latent codes $(\bm{y}_{\mathcal{A}},\bm{y}_{\mathcal{A}^c})$ and the contextual JSCC encoder parameters $(\bm{\phi}_{f_{a}},\bm{\phi}_{f_{na}})$, which are marked as the gradient descent updating terms ``$\cancelto{}{\Box}$'' in Fig. \ref{Fig_ntsccpp_system_architecture},
\begin{equation}\label{eq_instant_context_loss_func}
  \begin{aligned}
  & \mathcal{L}_{\text{RD}}(\bm{y}_{\mathcal{A}},\bm{y}_{\mathcal{A}^c},\bm{\phi}_{f_{a}},\bm{\phi}_{f_{na}}) = \\
  &  \mathbb{E}_{\bm{h}\sim {p_{\bm{h}}}} \big( -{\eta} \log{p_{{\bm{y}}|\boldsymbol{ z}}({\bm{y}}|\boldsymbol{ z})} + \lambda \cdot {d(\boldsymbol{x},\boldsymbol{\hat{x}})} \big),
\end{aligned}
\end{equation}
the other parameters $(\bm{\phi}_g,\bm{\phi}_h,\bm{\theta}_g,\bm{\theta}_f,\bm{\theta}_f)$ are amortized over the baseline contextual NTSCC model trained under the objective function in \eqref{eq_expect_context_loss_func}. A special note is that the expectation over the channel response in \eqref{eq_instant_context_loss_func} can be omitted if the instant channel response $\bm{h}$ can be known by the transmitter.

\subsection{Online Adaptation for ROI-based NTSCC++}

The variational estimate entropy on the latent codes represents the source content complexity, which guides the resource allocation in variable-rate JSCC. Apart from the entropy, the need of ROI-based coding stems from the fact that different pixels in an image have different levels of importance for human perception or machine tasks. It can be viewed as an extra semantic guidance to derive a new end-to-end RD loss function so that the spatially different channel bandwidth cost should be assigned according to ROI during coded transmission. Take the image source as example, when performing ROI-based NTSCC++, a quality map $\bm{m} \in \mathbb{R}^{H\times W}$ is assigned to the source image $\bm{x} \in \mathbb{R}^{H\times W \times 3}$, which indicates the semantic importance of each pixel. The ROI quality map $\bm{m}$ is bounded within $[0,1]$, where $0$ and $1$ represent the least and the most important pixel respectively. Guided by the ROI map $\bm{m}$ and its corresponding unbalanced bandwidth scaling factors $\{ \eta_i\}$ which are tied with the requirements from the low-level human vision or the high-level machine vision tasks \cite{song2021variable}, we extend the objective function in \eqref{eq_expect_context_loss_func} for obtaining the instance-adaptive target $\mathcal{L}_{\text{RD}}^{\text{ROI}}(\bm{y}_{\mathcal{A}},\bm{y}_{\mathcal{A}^c},\bm{\phi}_{f_{a}},\bm{\phi}_{f_{na}})$ as
\begin{equation}\label{eq_roi_instant_context_loss_func}
  \begin{aligned}
  & \mathcal{L}_{\text{RD}}^{\text{ROI}}(\bm{y}_{\mathcal{A}},\bm{y}_{\mathcal{A}^c},\bm{\phi}_{f_{a}},\bm{\phi}_{f_{na}}) = \\
  &  \mathbb{E}_{\bm{h}\sim {p_{\bm{h}}}} \big( \underbrace{\sum_{i}-{\eta_i} \log{p_{{ y_i}|\boldsymbol{ z}}({ y_i}|\boldsymbol{ z})}}_{\text{channel bandwidth cost}} + \lambda \cdot \underbrace{(\bm{m}\odot {d(\boldsymbol{x},\boldsymbol{\hat{x}})})}_{\text{ROI weighted distortion}} \big).
\end{aligned}
\end{equation}
Since our online instance-adaptive approach imposes no prior on the shape of ROI during model training, and thus our ROI control is quite flexible.

\subsection{Online Adaptation for Multi-Type Distortions}

For end-to-end wireless data transmission systems, sometimes we want to optimize the objective distortion (such as the mean-square-error (MSE)), sometimes we want to optimize the subjective perceptual loss (such as LPIPS \cite{lpips}), and sometimes we are more inclined to a balanced tradeoff between them \cite{ding2021comparison}. Since different distortion metrics are in odds to each other \cite{blau2019rethinking}, an intuitive solution is adding multiple-type distortions to \eqref{eq_expect_context_loss_func} to optimize NTSCC++ parameters for a specific weights between rate, and various distortions. However, multi-type distortion tradeoff requires multiple, or even infinite number of decoders to achieve, and a switch among different distortion weights may also need model retraining.

Again, we employ the online latent feature editing and JSCC encoder updating strategy to achieve the multi-distortion (MD) tradeoff by extending the NTSCC+ objective function in \eqref{eq_expect_context_loss_func}, we thus obtain the instance-adaptive target
\begin{equation}\label{eq_multi_instant_context_loss_func}
  \begin{aligned}
  & \mathcal{L}_{\text{RD}}^{\text{MD}}(\bm{y}_{\mathcal{A}},\bm{y}_{\mathcal{A}^c},\bm{\phi}_{f_{a}},\bm{\phi}_{f_{na}}) = \mathbb{E}_{\bm{h}\sim {p_{\bm{h}}}} \\
  &  \big( \underbrace{-{\eta} \log{p_{\boldsymbol{ y}|\boldsymbol{ z}}(\boldsymbol{ y}|\boldsymbol{ z})}}_{\text{channel bandwidth cost}} + \lambda_{o} \cdot \underbrace{{d_{\text{MSE}}(\boldsymbol{x},\boldsymbol{\hat{x}}})}_{\text{objective distortion}} + \lambda_{s} \cdot \underbrace{{d_{\text{LPIPS}}(\boldsymbol{x},\boldsymbol{\hat{x}}})}_{\text{subjective distortion}} \big),
\end{aligned}
\end{equation}
where the hyperparameters $\lambda_o$ and $\lambda_s$ control the tradeoff between the objective loss and the subjective loss. As marked in \eqref{eq_multi_instant_context_loss_func}, in this paper, these two types of loss are typically set as MSE and LPIPS, respectively, but we do not restrict other choices of objective and subjective distortions.

\begin{figure*}[t]
\setlength{\abovecaptionskip}{0.cm}
\setlength{\belowcaptionskip}{-0.cm}
	\centering{\includegraphics[width=\textwidth]{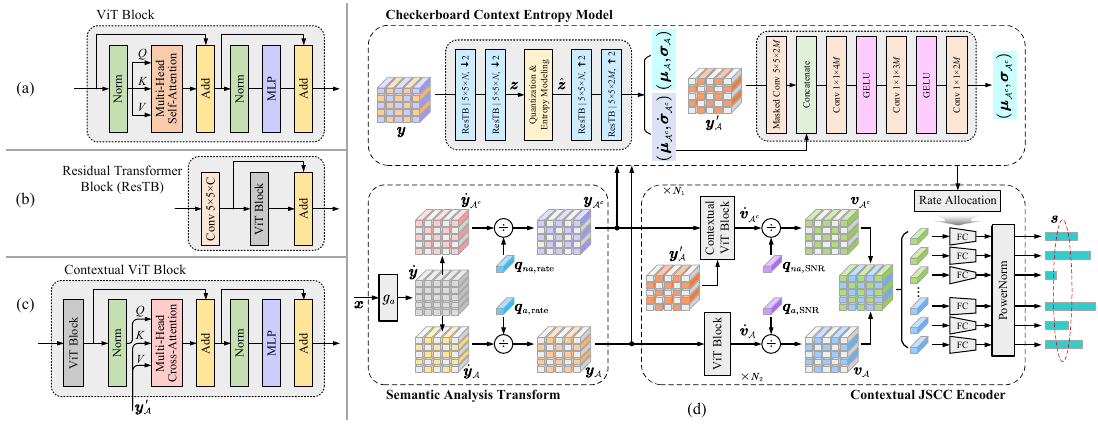}}
	\caption{The employed network architecture of the transmitter side of improved NTSCC for image sources. (a) shows the adopted vision Transformer (ViT) architecture. By integrating the ViT block, (b) shows the residual Transformer block (ResTB), (c) shows the contextual ViT block. Based on these blocks, we construct the transmitter of compatible NTSCC+ system in the subfigure (d), and the receiver is dual to this architecture. Convolution parameters are denoted as: ``the number of filters $\times$ kernel height $\times$ kernel width, up-sampling or down-sampling stride'', where $\uparrow$ / $\downarrow$ indicates up- and down-sampling, respectively.}\label{Fig_network_architecture}
	\vspace{-1em}
\end{figure*}

\section{Experimental Results}\label{section_results}

\subsection{Network Architecture}

The modular implementation details of our improved NTSCC are depicted in Fig. \ref{Fig_network_architecture}, where (a)(b)(c) present the structure of basic building blocks, and (d) shows the network architecture of the transmitter side.

\subsubsection{Nonlinear Transform and Entropy Model}

Build upon our previous work of NTSCC \cite{dai2022nonlinear}, we first modify the architecture of nonlinear transform modules $g_a^*$ and $g_s^*$ to build more expressive transforms. In particular, we replace the Transformer block using the residual Transformer Block (ResTB) \cite{liang2021swinir}, and incorporate a stride 2 convolution (transpose convolution) layer for downsampling (upsampling). Additionally, we utilize the checkerboard entropy model for more precise entropy estimation as shown at the top of Fig. \ref{Fig_network_architecture}(d). In this model, the information density of each patch is estimated using a parametric Gaussian density model.

In practice, by encapsulating a down-sampling layer and the following ResTBs as one stage, the analysis transform $g_a^*$ consists of 4 stages. We adopt $N=192$ feature channels for the first 3 stages and $M=320$ channels for the last stage. The numbers of ResTBs in four stages are set to (1, 2, 6, 2). On the other hand, the synthesis transform, $g_s^*$, has a symmetrical structure, with $M$ channels for the first stage, $N$ channels for the following 3 stages. The number of ResTBs in each stage is (2, 6, 2, 1). The multi-head self-attention (MHSA) within the ResTB employs a windows-based attention mechanism \cite{hassani2022neighborhood}, with a $7 \times 7$ sliding window for $g_a^*$ and $g_s^*$, and a $3 \times 3$ window for the entropy model. The number of heads in the MHSA increases by one for every 16 channels, and the channel expansion ratio of the multi-layer perceptron (MLP) is 2.

\subsubsection{Contextual JSCC Codec}

The contextual JSCC codec $f_e^*$ consists of two sub-encoders: the anchor branch $f_{e_a}$ and the non-anchor branch $f_{e_{na}}$, which are designed to exploit the contextual dependencies within JSCC codewords. To achieve this, we build a powerful contextual vision Transformer (ViT) block by conditioning $\bm{{y}}_{\mathcal{A}^c}$ on already decoded (simulated locally in the transmitter) semantic latent features $\bm{{y}}_{\mathcal{A}}^\prime$. This block (see Fig. \ref{Fig_network_architecture}(c)) alternates between self-attention layers and cross-attention layers. In the cross-attention layers, the anchor JSCC codewords $\bm{{y}}_{\mathcal{A}}^\prime$ are used to generate keys (K) and values (V), which will be queried by the latent representations in the non-anchor branch. We use $N_1 = 4$ contextual ViT blocks, and $N_2=4$ ViT blocks.

The JSCC rate matching process in the JSCC encoder $f_e^*$ maps the latent features $\bm{y}$ into variable-length channel-input vectors $\bm{s}$. The length of each channel-input vector $s_i$ is determined by a scalar quantization function $Q$, which is formulated as a piecewise linear function. Specifically, when $k_i\leq32$, the quantization is computed as $Q(k_i)=\min{(1, \lfloor k_i / 4 \rfloor)}$, whereas for $k_i>32$, the quantization is given by $Q(k_i)=\lfloor k_i / 16 \rfloor$. The resulting quantized rate allocation map is then reliably transmitted to the receiver as additional side information via a digital link using the portable network graphics (PNG) for source coding and LDPC codes for channel coding. Empirical results show that the cost of this side information ranges from $1\% \sim 7\%$ of the total bandwidth, depending on the specific CQI and SQI.

\subsection{Experimental Setup}

\subsubsection{Training Details}

DIV2K dataset \cite{agustsson2017ntire} is adopted as the training dataset. During training, images are randomly cropped into $256 \times 256$ patches. We use the Adam optimizer with default parameters. The training process includes a total of 610,000 iterations with the batchsize 8, where the learning rate is set to $10^{-4}$ in initial 600,000 iterations and decays to $10^{-5}$ for another 10,000 iterations. For the contextual NTSCC (NTSCC+), we train 5 models using 5 different $\lambda$ values in the set $\Lambda = \{0.013, 0.0483, 0.18, 0.36, 0.72\}$ and $\eta=0.2$, over the AWGN channel at SNR$=10$dB. For our compatible NTSCC+ scheme, we finetune the $\lambda=0.72$ model for extra 100,000 iterations. In each iteration, we randomly select $\lambda$ uniformly from $\Lambda$, draw $\eta$ values from $[0.15, 0.3]$, and sample channel SNR uniformly from $[0\text{dB}, 14\text{dB}]$. We set a group of global CQI scaling scalars $c_{\rm{global},\nu}$ with an interval of 3dB. Since the learned global scaling scalars only point to a few discrete $\lambda$ and $\nu$ values, the exponential interpolation is adopted between two adjacent scalars to achieve continuous rate-SNR adaptation. In the case of online instance-adaptive NTSCC++, without additional clarification, we update the latent codes $\bm{y}$ and the contextual JSCC encoder parameters $\bm{\phi}_f$ for $T_{\text{max}} = 20$ steps per image sample by default. The learning rate is set to $10^{-4}$ for $\bm{\phi}_f$ and $5 \times 10^{-3}$ for $\bm{y}$. The best updated codes and model parameters based on the RD loss is selected for model inference.

\subsubsection{Evaluation Protocol}

We evaluate the image transmission systems on two popular datasets that contain diverse contents and resolutions: Kodak \cite{Kodak} at the size of $512 \times 768$, and CLIC21 testset \cite{CLIC21} with approximate 2K resolution ($2560 \times 1440$) images. These datasets are widely used for image processing evaluation. We mainly consider the mean square error (MSE) for distortion measurement calculated in RGB color space, which is aligned with the optimization target of the classical image codec. We compare our method with a broad collection of prominent image codecs combined with 5G LDPC channel coding \cite{richardson2018design} and digital modulation following the 3GPP standard. In particular, the considered conventional transform-based image codecs include BPG (compliant with the intra-frame coding scheme of the HEVC standard) and VTM-20.0 (intra-frame coding scheme of the VVC standard, SOTA engineered image codec). We also consider the emerging end-to-end learned image compression methods incudes Ball{\'e} \emph{et al.} 2018 \cite{balle2018}, Minnen \emph{et al.} 2018 \cite{minnen2018}, and He \emph{et al.} 2021 \cite{he2021checkerboard} for comparison. Apart from these, we also compare our improved NTSCC with the deep JSCC scheme \cite{djscc} and the original NTSCC \cite{dai2022nonlinear} for end-to-end transmission. For fair comparison, all images are cropped to multiples of 64 to avoid padding for neural codecs. The above simulations are implemented at the top of Sionna \cite{hoydis2022sionna}, which is an open-source library for link-level simulations of digital communication systems. We use ``+'' to concatenate the source coding and channel coding schemes. Please refer to the Appendix \ref*{appendix_a} for more detailed experiment settings.

\subsection{Results Analysis}

\begin{figure*}[t]
\setlength{\abovecaptionskip}{0.cm}
\setlength{\belowcaptionskip}{-0.cm}
 \centering
 \subfigure[Kodak]{\includegraphics[height=0.13\textheight]{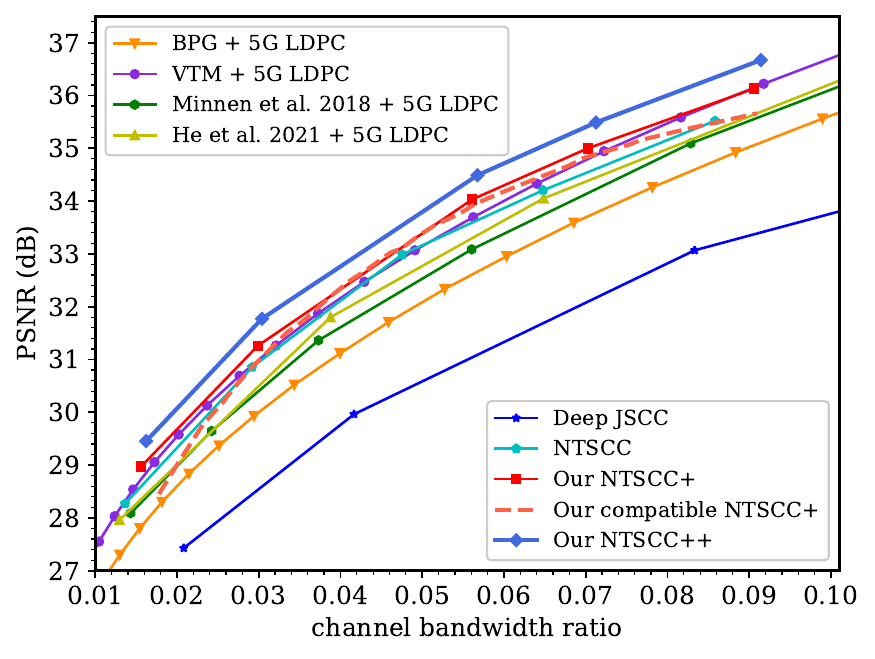}}
 \subfigure[Kodak]{\includegraphics[height=0.132\textheight]{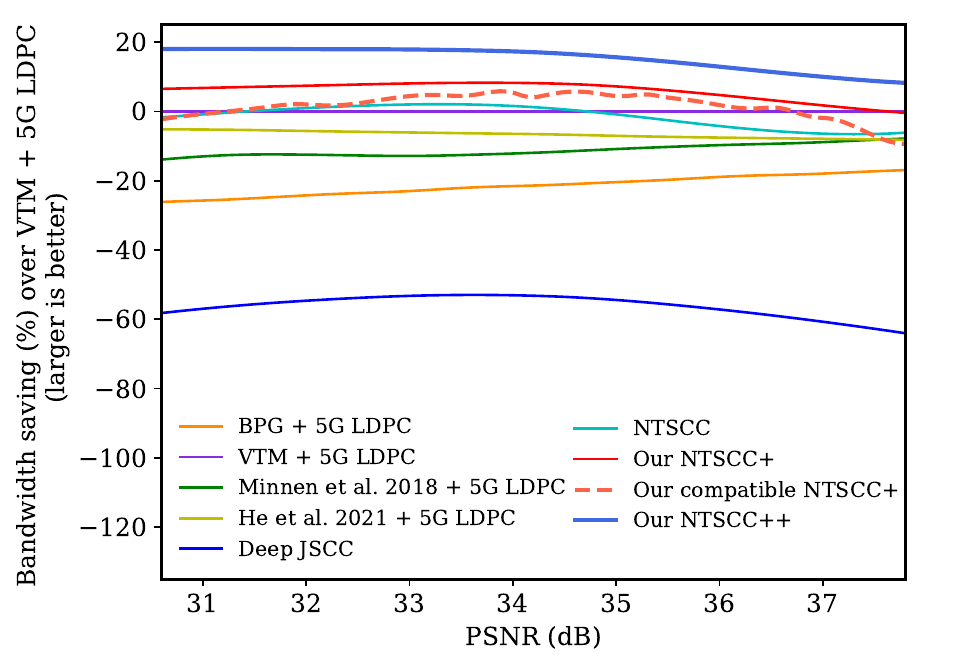}}
 \subfigure[CLIC21]{\includegraphics[height=0.13\textheight]{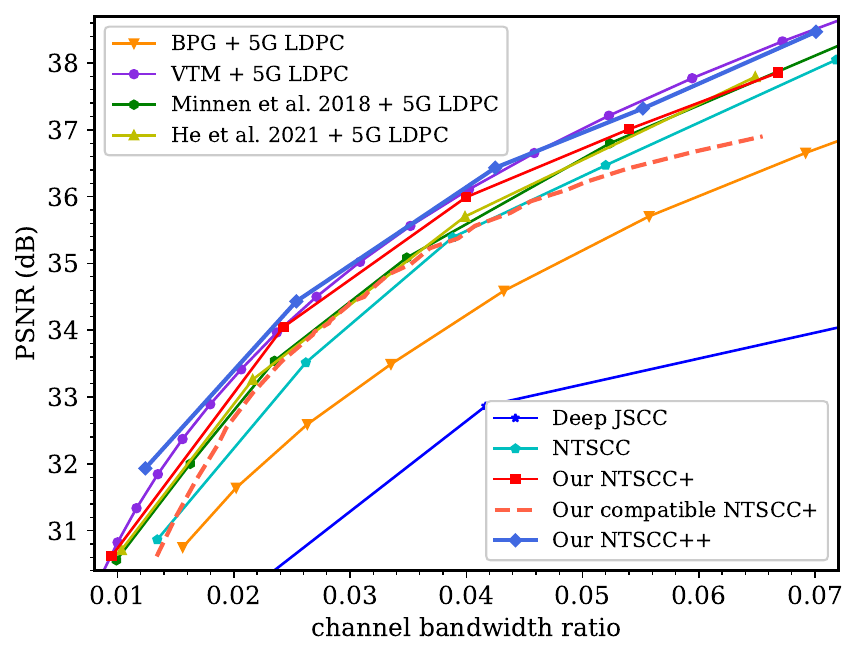}}
 \subfigure[CLIC21]{\includegraphics[height=0.132\textheight]{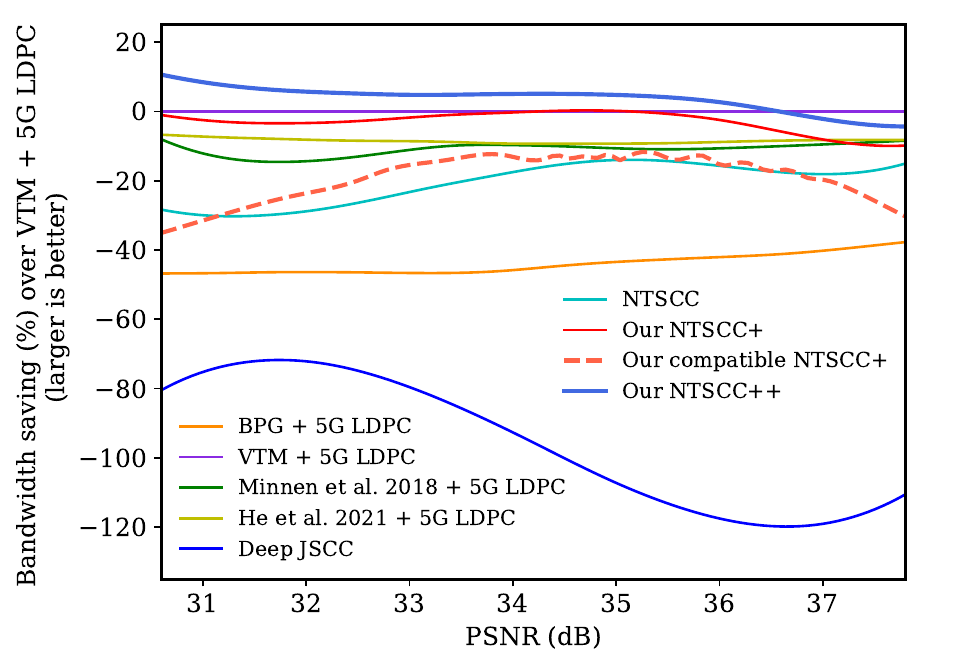}}
 \caption{End-to-end rate-distortion comparison over the AWGN channel at $\text{SNR} = 10$dB for (a) Kodak (c) CLIC21. Bandwidth saving against the ``VTM + 5G LDPC''
 		anchor for (b) Kodak (d) CLIC21. Peak signal-to-noise ratio (PSNR) is used to measure reconstruction quality.}
 \label{Fig_rd_curve}
\end{figure*}

\begin{figure*}[t]
\setlength{\abovecaptionskip}{0.cm}
\setlength{\belowcaptionskip}{-0.cm}
 \centering
 \subfigure[]{\includegraphics[height=0.17\textheight]{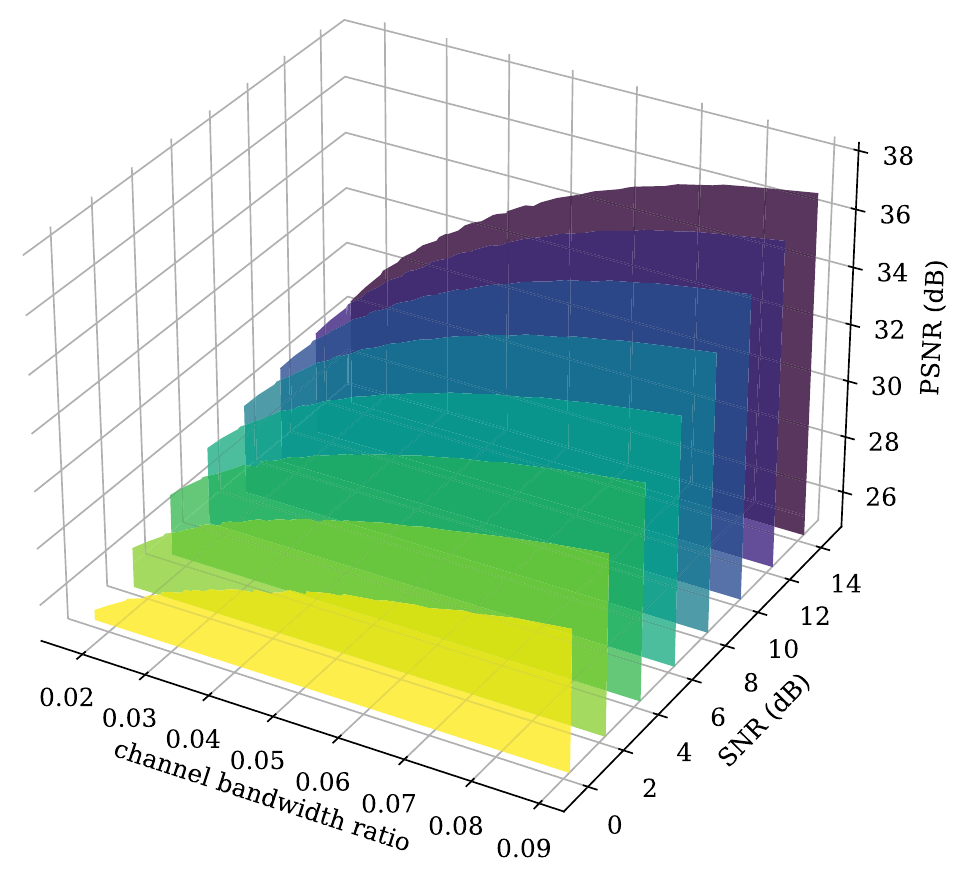}}
 \hspace{1em}
 \subfigure[]{\includegraphics[height=0.17\textheight]{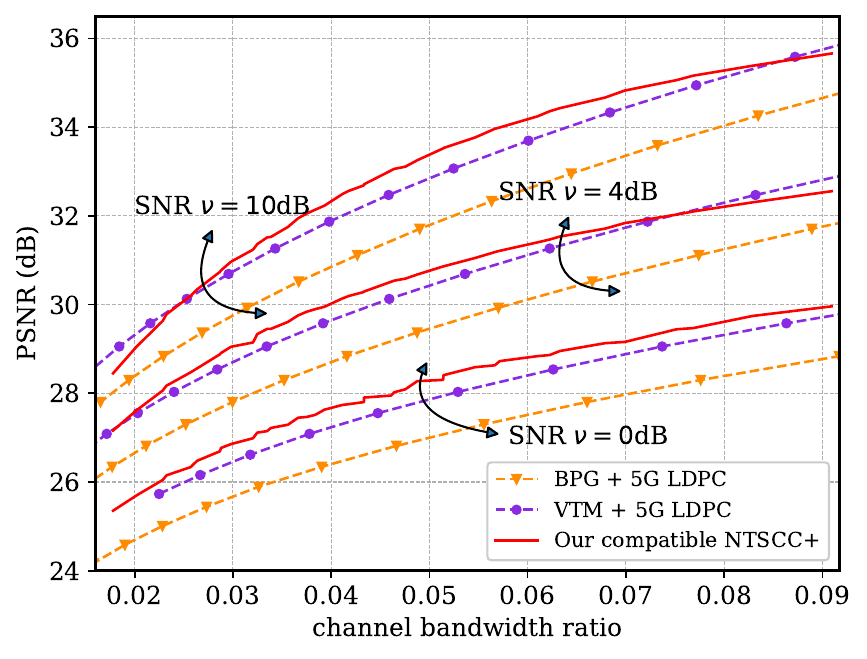}}
 \hspace{1em}
 \subfigure[]{\includegraphics[height=0.17\textheight]{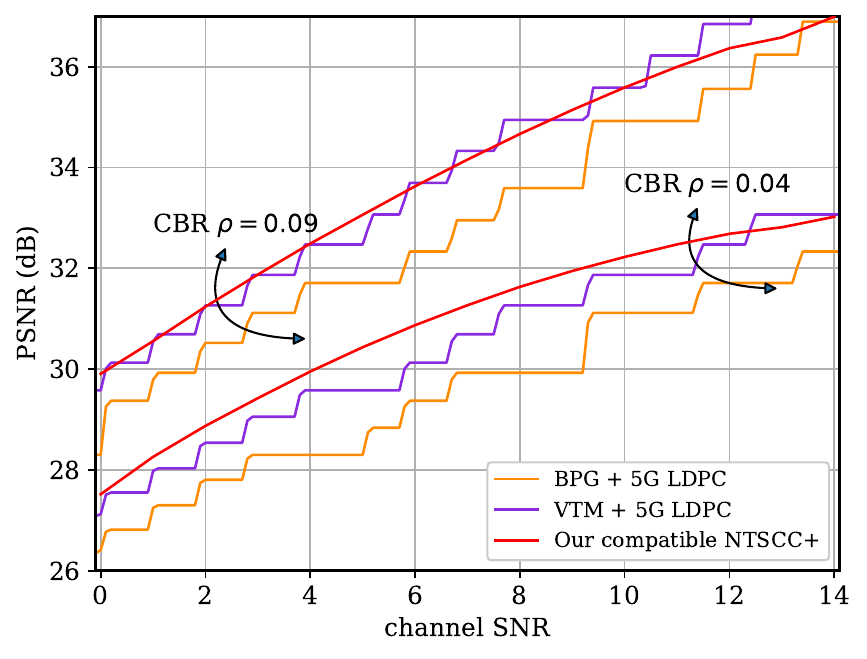}}
 \caption{End-to-end rate-SNR-distortion comparison over the AWGN channel on the Kodak dataset. (a) shows the rate-SNR-distortion surface obtained by compatible NTSCC+. (b) compares RD curves of different coded transmission schemes at SNR $= 0$dB, $4$dB, and $10$dB. (c) compares SNR-distortion curves under the CBR constraint $\rho=0.04$ and $\rho=0.09$.}
 \label{Fig_versatile_results}
\end{figure*}

Fig. \ref{Fig_rd_curve} shows the RD comparison results over the AWGN channel at SNR $=10$dB. These results indicate that our contextual NTSCC model (NTSCC+) can outperform all of the traditional coded transmission schemes using the standardized image codec (BPG and VTM) or the emerging neural image codecs. To the best of our knowledge, this is the first end-to-end transmission system outperforming the SOTA VTM + 5G LDPC scheme on the classical PSNR metric. Compared to the standard NTSCC, our NTSCC+ shows clear performance gain, that verifies the efficiency of checkerboard context model on exploiting the correlative probabilistic structure among the semantic latent features. Moreover, through online updating latents and JSCC encoder, our NTSCC++ achieves further performance gain in Fig. \ref{Fig_rd_curve}(a) with some extra encoding complexity and latency cost as discussed later. In Fig. \ref{Fig_rd_curve}(b), we present the relative CBR reduction against the ``VTM + 5G LDPC'' anchor at a range of PSNR levels. These results are derived according to the Bj\o{}ntegaard Delta (BD) chart \cite{bjontegaard2001calculation} by plotting the CBR savings as a function of quality, which verifies that the system coding efficiency gain of our NTSCC+ and NTSCC++ holds for a wide PSNR range and diverse image resolutions. For completeness, we present more RD curves and results tested on Kodak and other datasets in the Appendix \ref*{appendix_b}.

In Table \ref{table_bd_rate}, we summarize the BD-rate (smaller is better) of coded transmission schemes across two datasets. Compared to the He \emph{et al.} 2021 \cite{he2021checkerboard} + 5G LDPC which also involves attention-based transforms and checkerboard context model, our NTSCC+ still presents superiority, that justifies the effectiveness of our well-matched nonlinear transform, context entropy modeling, and contextual JSCC codec. Compared to the anchor VTM + 5G LDPC, our NTSCC++ achieves a 14.25\% channel bandwidth cost saving on Kodak dataset and a 4.70\% CBR saving on CLIC21 testset.

\begin{table}[t]
	
	\centering
	\renewcommand{\arraystretch}{1.3}
	
	\caption{Averaged BD-rate (smaller is better) improvement against the ``VTM + 5G LDPC'' anchor for different transmission schemes and datasets. \textcolor{red}{Red} and \textcolor{blue}{blue} indicates the best and the second best performance, respectively.}
	
	\begin{tabular}{m{4.3cm}|m{1.6cm}|m{1.6cm}}
		
		\Xhline{1pt}
		
		{Coded transmission scheme} & \centering \shortstack{Kodak} & \centering \shortstack{CLIC21} \tabularnewline
		\hline
		\hline
		JPEG + 5G LDPC &\centering 186.09\% & \centering 586.31\% \tabularnewline
		
		JPEG2000 + 5G LDPC &\centering 94.24\% & \centering 323.11\% \tabularnewline
		
		BPG + 5G LDPC &\centering 23.93\% & \centering 55.43\% \tabularnewline
		
		VTM + 5G LDPC &\centering  \textbf{0\%} & \centering \textbf{0\%} \tabularnewline
		
		\hline
		
		Ball{\'e} \emph{et al.}\cite{balle2018} + 5G LDPC &\centering  31.79\%  & \centering 47.71\% \tabularnewline
		
		Minnen \emph{et al.}\cite{minnen2018} + 5G LDPC  &\centering 11.94\%  & \centering 11.65\% \tabularnewline
		
		He \emph{et al.}\cite{he2021checkerboard} + 5G LDPC  &\centering  6.86\% & \centering 8.90\% \tabularnewline
		
		\hline
		
		Deep JSCC &\centering 74.87\% & \centering 158.88\% \tabularnewline
		
		NTSCC  &\centering  1.33\% & \centering 23.06\% \tabularnewline
		
		\textbf{Our NTSCC+} &\centering \textbf{\textcolor{blue}{--5.86\%}} & \centering \textbf{\textcolor{blue}{2.94\%}} \tabularnewline
		
		\textbf{Our compatible NTSCC+} &\centering --1.74\% & \centering 20.94\% \tabularnewline
		
		\textbf{Our NTSCC++} &\centering \textbf{\textcolor{red}{--14.25\%}} & \centering \textbf{\textcolor{red}{--4.70\%}} \tabularnewline
		
		\Xhline{1pt}
	\end{tabular}
	\label{table_bd_rate}
\end{table}

\makeatletter
\renewcommand{\@thesubfigure}{\hskip\subfiglabelskip}
\makeatother

\begin{figure*}[t]
	\setlength{\abovecaptionskip}{0.cm}
	\setlength{\belowcaptionskip}{-0.cm}
	\begin{subtable}
		\centering
		\small
		\begin{tabular}{m{0.235\textwidth}<{\centering}m{0.22\textwidth}<{\centering}m{0.22\textwidth}<{\centering}m{0.22\textwidth}<{\centering}}
			Original image / ROI map & VTM + 5G LDPC & \shortstack{Our standard NTSCC++} &  \shortstack{Our ROI-based NTSCC++}
		\end{tabular}
	\end{subtable}
	
	\vspace{-0.5em}
	
	\begin{center}
		\hspace{-.05in}
		\subfigure[] {\includegraphics[width=0.235\textwidth]{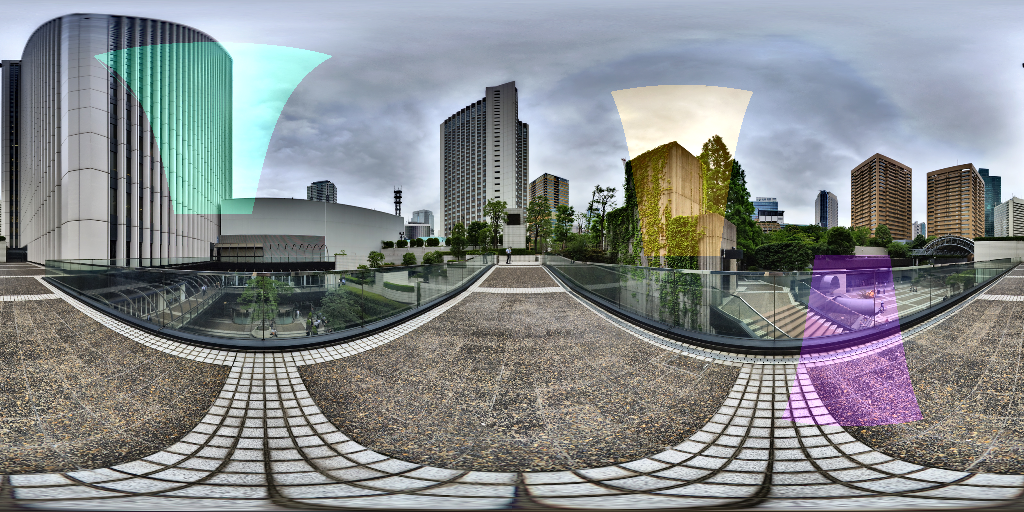}}
		\hspace{-.15in}
		\quad
		\subfigure[] {\includegraphics[width=0.235\textwidth]{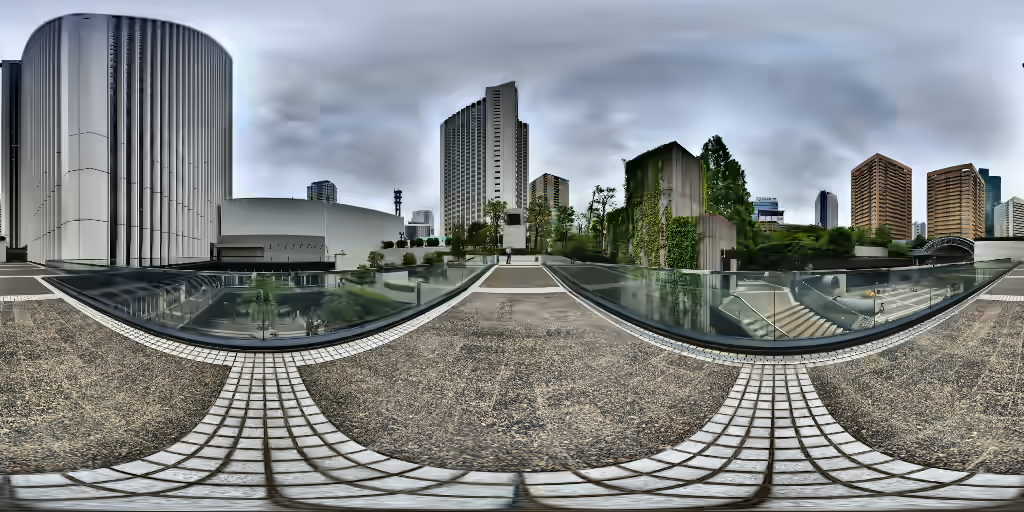}}
		\hspace{-.15in}
		\quad
		\subfigure[] {\includegraphics[width=0.235\textwidth]{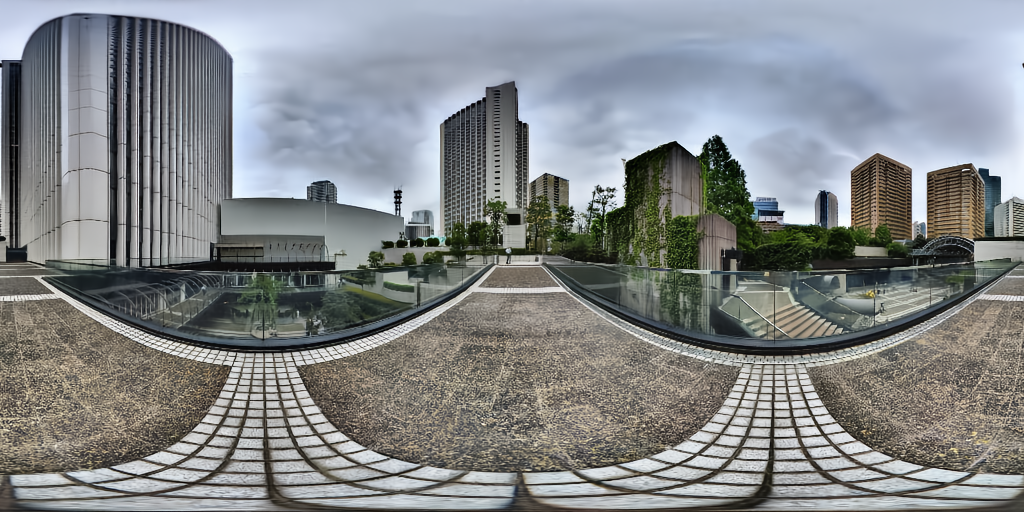}}
		\hspace{-.15in}
		\quad
		\subfigure[] {\includegraphics[width=0.235\textwidth]{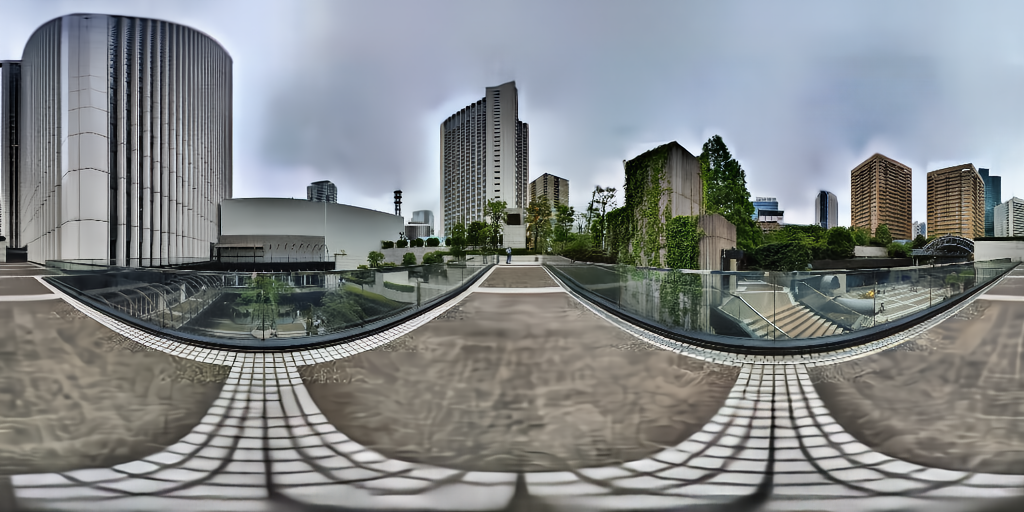}}
		
		\vspace{-.20in}
		\hspace{-.05in}
		\subfigure[$\rho$ / PSNR / PSNR$_{\text{ROI}}$] {\includegraphics[width=0.235\textwidth]{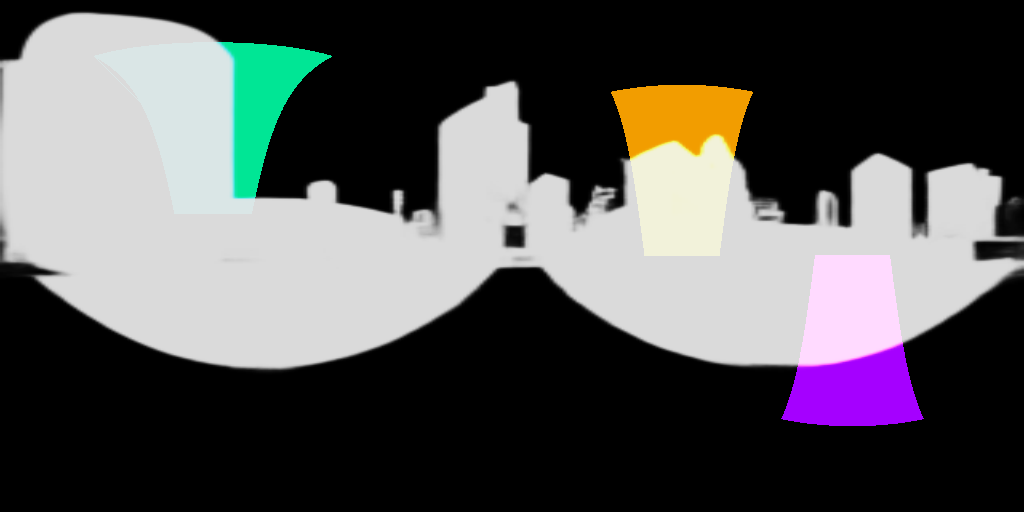}}
		\hspace{-.15in}
		\quad
		\subfigure[\textbf{0.062} / 26.9 / 26.6] {\includegraphics[width=0.235\textwidth]{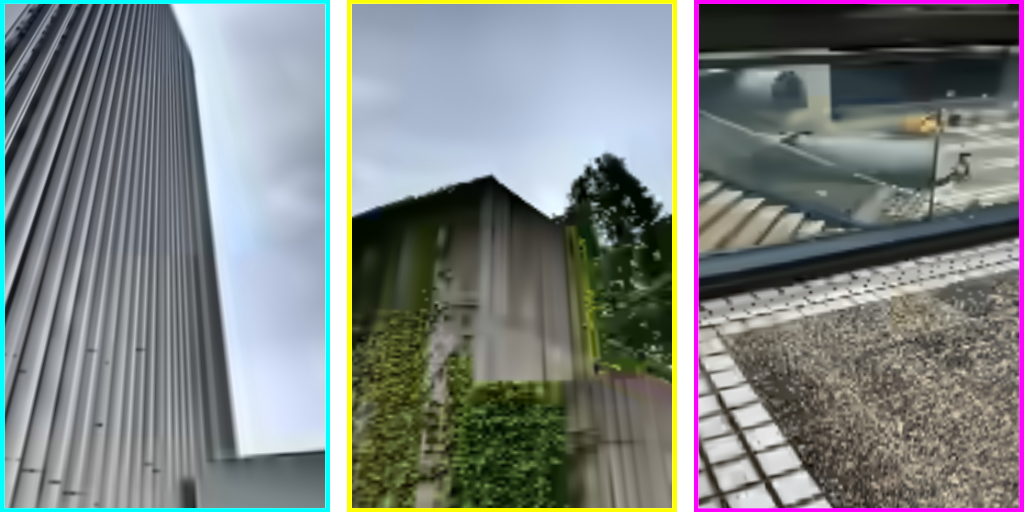}}
		\hspace{-.15in}
		\quad
		\subfigure[\textbf{0.062} / 27.0 / 27.5] {\includegraphics[width=0.235\textwidth]{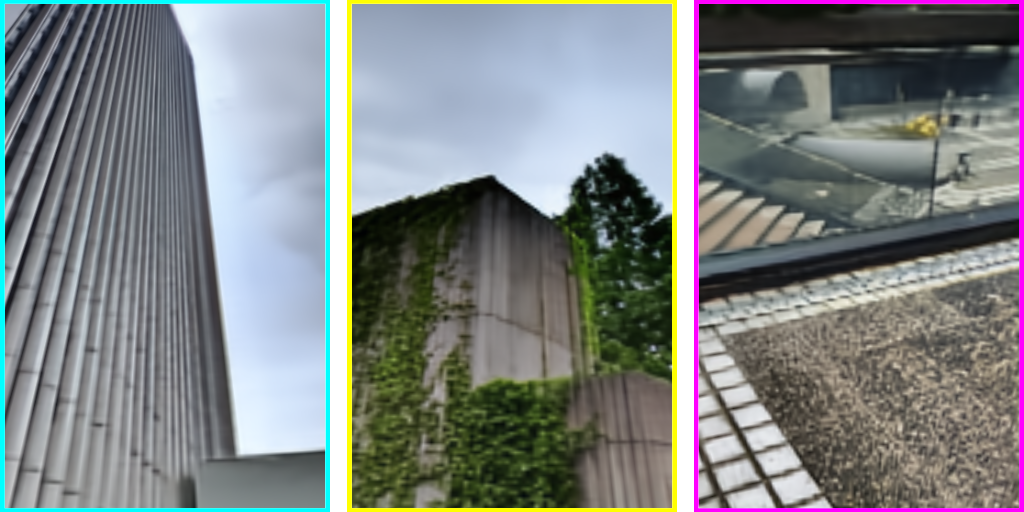}}
		\hspace{-.15in}
		\quad
		\subfigure[\textbf{0.038} / 21.8 / 27.6] {\includegraphics[width=0.235\textwidth]{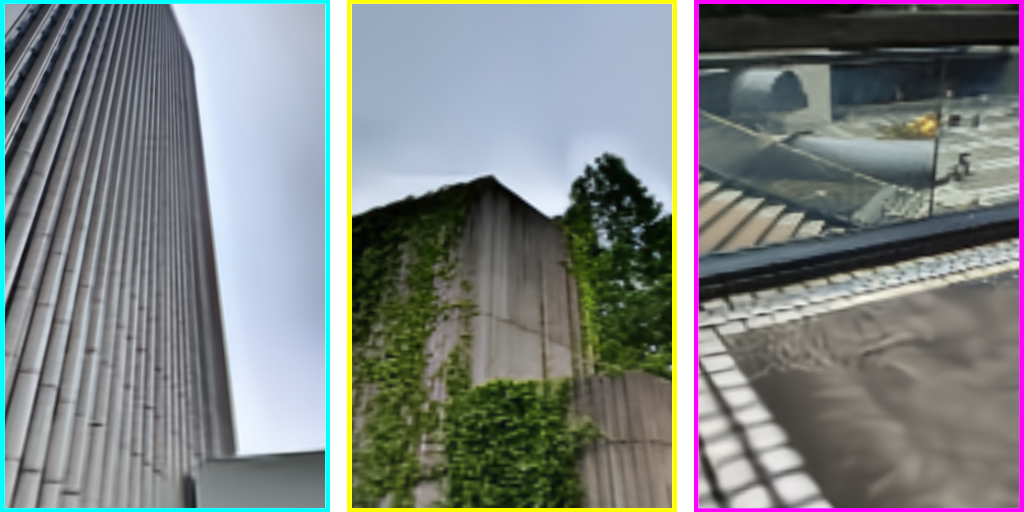}}
		
		\caption{Visual examples of online adaptation based on the ROI over the AWGN channel with a SNR of 10dB. The first column shows the original equirectangular projection (ERP) image and its corresponding ROI mask, in which 3 viewports with different visual angles are selected and distinguished by different colors. The subsequent columns display the reconstructed images generated by three coded transmission schemes. In the implementation of the ROI-based NTSCC++ scheme, the scaling factor ${\eta}_1 = 0.25$ and the quality map value ${m}_1 = 1.25$ are assigned to the ROI regions, indicated by the white color in the ROI mask, while the scaling factor ${\eta}_2 = 0.05$ and the quality map value $m_2 = 0.25$ are assigned to the non-ROI regions, indicated by the black color in the ROI mask.}
		\label{Fig_visual_roi}
	\end{center}
	\vspace{-1em}
\end{figure*}

\begin{figure}[t]
\setlength{\abovecaptionskip}{0.cm}
\setlength{\belowcaptionskip}{-0.cm}
 \centering
 {\includegraphics[scale=0.44]{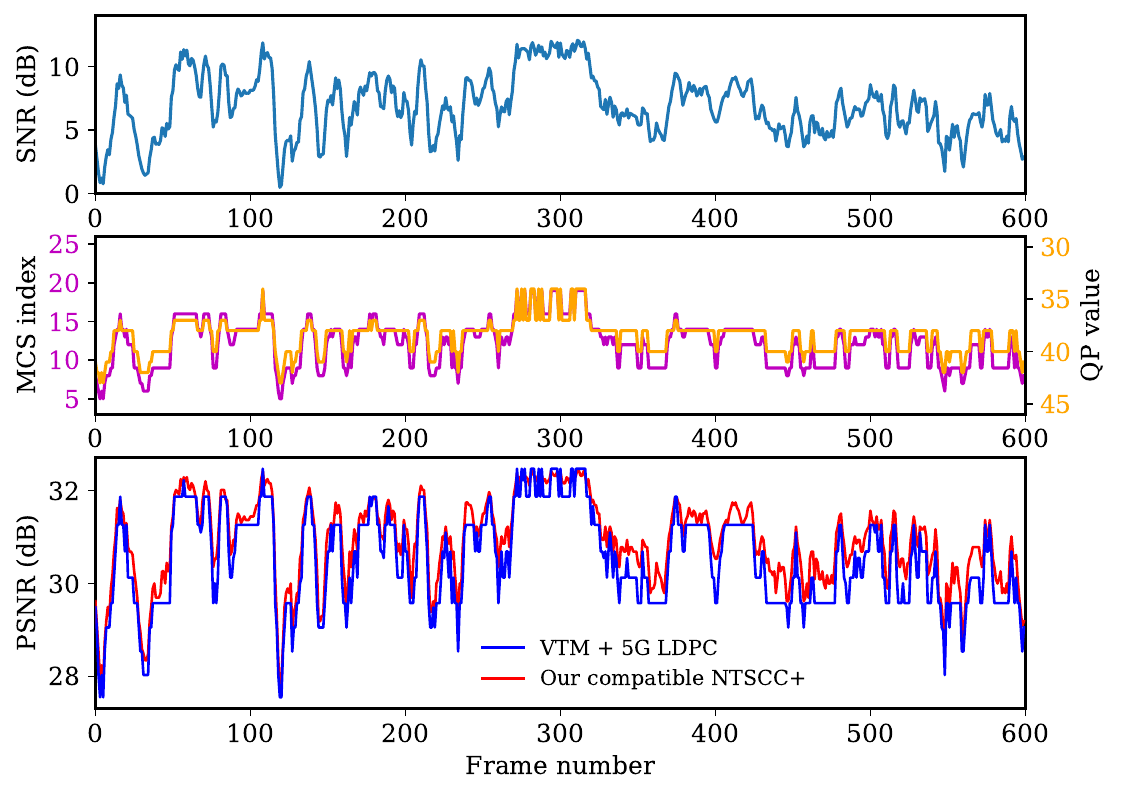}}
 \caption{Comparisons of the image quality between the VTM + 5G LDPC and our compatible NTSCC+ under a practical fading channel sampled from COST2100 channel model \cite{liu2012cost}, where the fading coefficient varies with the frame number. The top subfigure shows the instant channel SNR, the middle subfigure shows the adaptive coded modulation scheme and the quantization parameter (QP) in VTM + 5G LDPC. The bottom subfigure plots the PSNR value of each frame under the CBR constraint $\rho=0.04$.
 }
 \label{Fig_fading_results}
\end{figure}

In Fig. \ref{Fig_versatile_results}, we show the performance of our bandwidth compatible contextual NTSCC (compatible NTSCC+) over the AWGN channel with different SNR values and under various CBR constraint. Results in Fig. \ref{Fig_versatile_results}(a) indicate that our compatible NTSCC+ can achieve \emph{fine continuous} rate control and SNR adaptation in a single model, which can effectively reduce computational and storage cost during model training and deployment. Although compatible NTSCC+ performs worse than the multi-model NTSCC+, it is still competitive for better performance than BPG + 5G LDPC and on par performance with VTM + 5G LDPC over a broad SNR and CBR region.

To further evaluate the flexibility of our compatible NTSCC+, Fig. \ref{Fig_fading_results} illustrates a practical fading channel case collected from COST2100 channel model \cite{liu2012cost}. The results indicate that, the classical layered design, with finite number of QPs in source compression and modulation and coding schemes (MCS) for error protection, fails to reacts sensitive to fluctuations in SNR. Conversely, our compatible NTSCC+ with response networks is shown to not only provide smooth performance transition without any cliff and leveling effects but also achieve competitive performance.

Fig. \ref{Fig_visual_roi} shows a visual comparison among VTM + 5G LDPC, our standard instance-adaptive NTSCC++, and our ROI-based NTSCC++ on a 360$^\circ$ image viewed from 3 viewports, a typical source in XR. We can clearly see that our NTSCC++ maintains more details of high frequency components at the equivalent channel bandwidth rates. In contrast, VTM + 5G LDPC exhibits visual artifacts such as blocking, aliasing and ringing, especially near edges. Additionally, our ROI-based NTSCC++ method offers a significant advantage in terms of bandwidth savings. It achieves a 40\% bandwidth rate savings over the standard NTSCC++ method, while still preserving the quality of the images within the ROI. This highlights the potential of the method as a viable solution for 360$^\circ$ media applications such as viewport-based streaming or saliency-aware adaptive transmission, as discussed in \cite{xu360}. For more visualization results, please refer to Appendix \ref*{appendix_b}.

\begin{table}[t]
	\renewcommand{\arraystretch}{1.3}
	\centering
	
	\caption{Averaged encoding/decoding latency on the Kodak dataset. }
	
    \begin{threeparttable}
	\begin{tabular}{m{3.05cm}|m{0.8cm}|m{0.8cm}|m{1.0cm}|m{1.0cm}}
		
		\Xhline{1pt}
		
		\multirow{2}*{\shortstack{Transmission scheme}} & \centering \multirow{2}*{\shortstack{BD-$\rho$\\(\%, $\downarrow$)}} & \centering \multirow{2}*{\shortstack{Params\\(M)}} & \multicolumn{2}{c}{End-to-end latency (s)} \tabularnewline
		\cline{4-5}
		~ & ~ & ~ & \centering Encoding & \centering Decoding \tabularnewline
		\hline
		\hline

		Minnen \cite{minnen2018} + 5G LDPC  & \centering 11.94 & \centering 14.13 \tnote{\dag} &\centering  0.046 & \centering 1.7 \tabularnewline

		He \cite{he2021checkerboard} + 5G LDPC  &  \centering 6.86 & \centering 26.60 \tnote{\dag} & \centering 0.087 & \centering 0.15 \tabularnewline

		VTM + 5G LDPC  & \centering 0 & \centering --- & \centering 65.8 & \centering 0.289 \tabularnewline
		\hline
		
		NTSCC \cite{dai2022nonlinear} & \centering 1.33 & \centering 32.94 \tnote{\ddag} & \centering 0.036 & \centering 0.015 \tabularnewline		
		
		Our NTSCC+  & \centering --5.86 & \centering 57.36 \tnote{\ddag} & \centering 0.058 & \centering 0.023 \tabularnewline
		
		Our compatible NTSCC+ & \centering --1.74 & \centering 60.29 & \centering 0.061 & \centering 0.027 \tabularnewline
		
		Our NTSCC++ (1 step) & \centering --9.77 & \centering 57.36 \tnote{\ddag} & \centering 0.6 & \centering 0.023 \tabularnewline
				
		Our NTSCC++ (5 steps) & \centering --11.30 & \centering 57.36 \tnote{\ddag} & \centering 2.5 & \centering 0.023 \tabularnewline
		
		Our NTSCC++ (10 steps) & \centering --12.80 & \centering 57.36 \tnote{\ddag} & \centering 5.2 & \centering 0.023 \tabularnewline
		
		Our NTSCC++ (20 steps) & \centering --14.25 & \centering 57.36 \tnote{\ddag} & \centering 10.9 & \centering 0.023 \tabularnewline
		
		Our NTSCC++ (100 steps) & \centering --16.75 & \centering 57.36 \tnote{\ddag} & \centering 52.3 & \centering 0.023 \tabularnewline
		
		\Xhline{1pt}
	\end{tabular}
    \begin{tablenotes}
      \item[\dag] The total model size should be $N_{\lambda}$ times larger, where $N_{\lambda}$ is the number of rate points.
      \item[\ddag] The total model size should be $N_{\lambda} \times N_{\nu}$ times larger, where $N_{\nu}$ denotes the number of SNR points.
    \end{tablenotes}
    \end{threeparttable}
	\label{table_latency}
\vspace{-1em}
\end{table}

We evaluate the end-to-end processing latency of these coded transmission schemes on the Kodak dataset and show the metrics in Table \ref{table_latency}, including the BD-rate ($\rho$, ``$\downarrow$'' means smaller is better), single model parameters, and encoding (decoding) latency. All experiments are conducted using PyTorch 1.9.0, with the Inter Xeon Gold 6226R CPU (mainly for arithmetic coding) and one RTX 3090 GPU (for neural network inference and LDPC codec). Compared to other neural coded transmission schemes, our compatible NTSCC+ is more storage-efficient during deployment as it only requires a single model to perform both rate adjustment and channel state adaptation, whereas multiple models are necessary for other schemes to achieve these functionalities. It can also be observed that our NTSCC based end-to-end transmission schemes run clearly faster than classical layered designed schemes, attributed to the absence of arithmetic coding and LDPC decoding. Compared with the original NTSCC \cite{dai2022nonlinear}, the encoding/decoding time of our improved version increases slightly. However, it is valuable since the end-to-end transmission efficiency shows clear superiority (surpassing the SOTA VTM + 5G LDPC). In addition, by dedicating extra \emph{encoding time} to NTSCC++, even better rate-distortion performance can be achieved \emph{without increasing decoding time}. Our experiments indicate that only 20 online updating steps provide a reasonable trade-off between speed and performance, which can facilitate XR applications.

\section{Conclusion}\label{section_conclusion}

In this paper, from the perspectives of model accuracy, versatility, and adaptability, we have proposed a comprehensive framework for improving NTSCC to boost the semantic coded transmission. This new sophisticated improved NTSCC system is ready to support low-latency data interaction in XR, which can catalyze the application of semantic communications.

\ifCLASSOPTIONcaptionsoff
  \newpage
\fi

\bibliographystyle{IEEEtran}
\bibliography{Ref}

\begin{appendices}
	\section{Separation-based transmission system evaluation}
	\label{appendix_a}
	In this section, we provide the detailed experiment configurations used for separation-based transmission schemes. The settings of non-neural image codecs and LDPC codes are listed below in the following subsection. For the neural image codecs, we employ the pre-trained models from the CompressAI library \cite{begaint2020compressai}. 
	
	\subsection{VTM}
	We use the VVC test model software VTM-20.0 which is built from the website\footnote{https://vcgit.hhi.fraunhofer.de/jvet/VVCSoftware\_VTM/-/releases/VTM-20.0}. Since VVC operates on the YUV color space by default, but the test dataset we used are RGB format. We first convert RGB image to YUV 4:4:4 color space using OpenCV, and compress the YUV files with the command line
	\begin{minted}[
		frame=lines,
		framesep=2mm,
		baselinestretch=1.2,
		bgcolor=white,
		fontsize=\footnotesize,
		]{python}
# VTM encoding
EncoderAppStatic -c encoder_intra_vtm.cfg  
-q [QP] -i [original_yuv_file] 
-b [encoded_bitstream] -f 1 -fr 1 
-wdt [width] -hgt [height] 
--InputChromaFormat=444
	\end{minted}
	where \emph{encoder$\_$intra$\_$vtm.cfg} is the official intra configuration files by default. With reliable transmission of the bitstream, we can reconstruct source image in the YUV 4:4:4 format with the command line
	\begin{minted}[
		frame=lines,
		framesep=2mm,
		baselinestretch=1.2,
		bgcolor=white,
		fontsize=\footnotesize
		]{python}
# VTM decoding
DecoderAppStatic -d 8 -b [recovered_bitstream] 
-o [decoded_yuv_file] 	
	\end{minted}
	where the \emph{recovered$\_$bitstream} is the reconstruction bits from channel decoding. The reconstructed image has the YUV 4:4:4 format, and we converted it back to RGB space again using OpenCV. The final PSNR is computed between the final RGB image and the original RGB image.
	\subsection{BPG}
	BPG software is obtained from the website\footnote{https://bellard.org/bpg/} and the following commands are used for encoding and decoding.
	\begin{minted}[
		frame=lines,
		framesep=2mm,
		baselinestretch=1.2,
		bgcolor=white,
		fontsize=\footnotesize,
		]{python}
# BPG encoding
bpgenc -e x265 -q [QP] -f 444
-o [encoded_bitstream] [original_png_file]
# BPG decoding
bpgdec -o [decoded_png_file] [recovered_bitstream]
	\end{minted}

	\subsection{LDPC Codes}
	
	\begin{figure}[t]
		\setlength{\abovecaptionskip}{0.cm}
		\setlength{\belowcaptionskip}{-0.cm}
		\centering
		\subfigure[]{\includegraphics[scale=0.5]{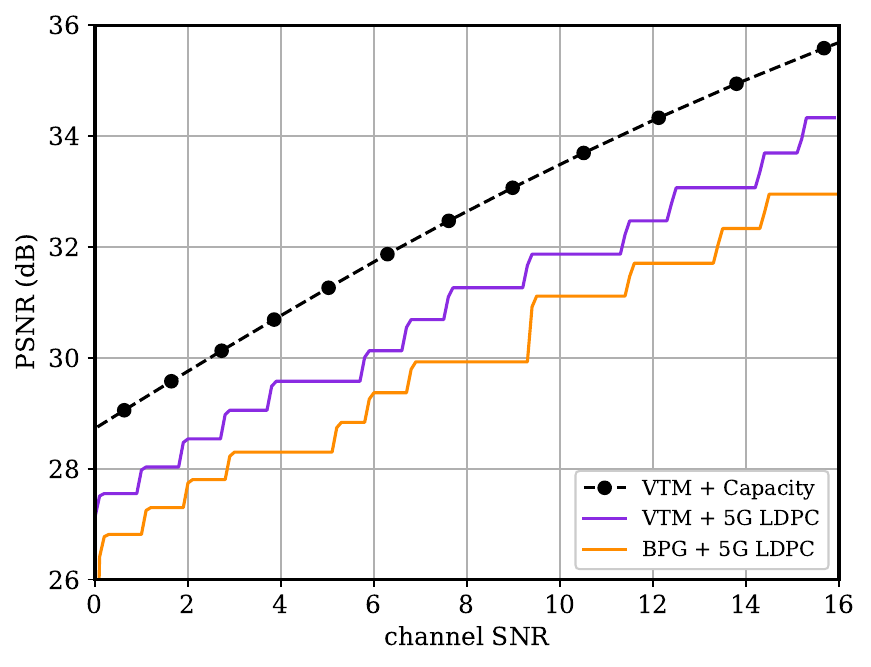}}
		\caption{The performance of LDPC codes used in this paper. ``VTM + Capacity'' means VTM combined with a capacity achieving channel code.}
		\label{Fig_vtm_ldpc}
		\vspace{-1em}
	\end{figure}
	
	We employ 5G LDPC codes \cite{richardson2018design} with length 4096 for error correction. The combinations of modulation order and channel code rate are selected according the modulation and coding scheme (MCS) Table 5.1.3.1-1 for physical downlink shared channel (PDSCH) in 3GPP TS 38.214 version 16.2.0\footnote{https://www.etsi.org/deliver/etsi\_ts/138200\_138299/138214/16.02.00\_60}. For a specific MCS index, we use the maximum source coding rate (the minimum QP value) within the CBR target. If the image decoding procedure fails due to transmitting errors in the compressed bitstream, we use the mean value for all the pixels per color channel as the reconstruction image. We select the best performing configurations in terms of average distortion after traversing all possible MCS choices at each channel SNR. The above simulations are implemented at the top of Sionna\footnote{https://github.com/NVlabs/sionna}. Fig. \ref{Fig_vtm_ldpc} presents the SNR-PSNR curve of VTM + 5G LDPC under AWGN channel with the CBR target $\rho=0.04$, where the capacity achieving bound ``VTM + Capacity'' also plotted for reference.
	\vspace{-1em}
	\section{Additional Results} \label{appendix_b}
	\begin{figure*}[t]
		\setlength{\abovecaptionskip}{0.cm}
		\setlength{\belowcaptionskip}{-0.cm}
		\centering
		\subfigure[]{\includegraphics[width=0.9\textwidth]{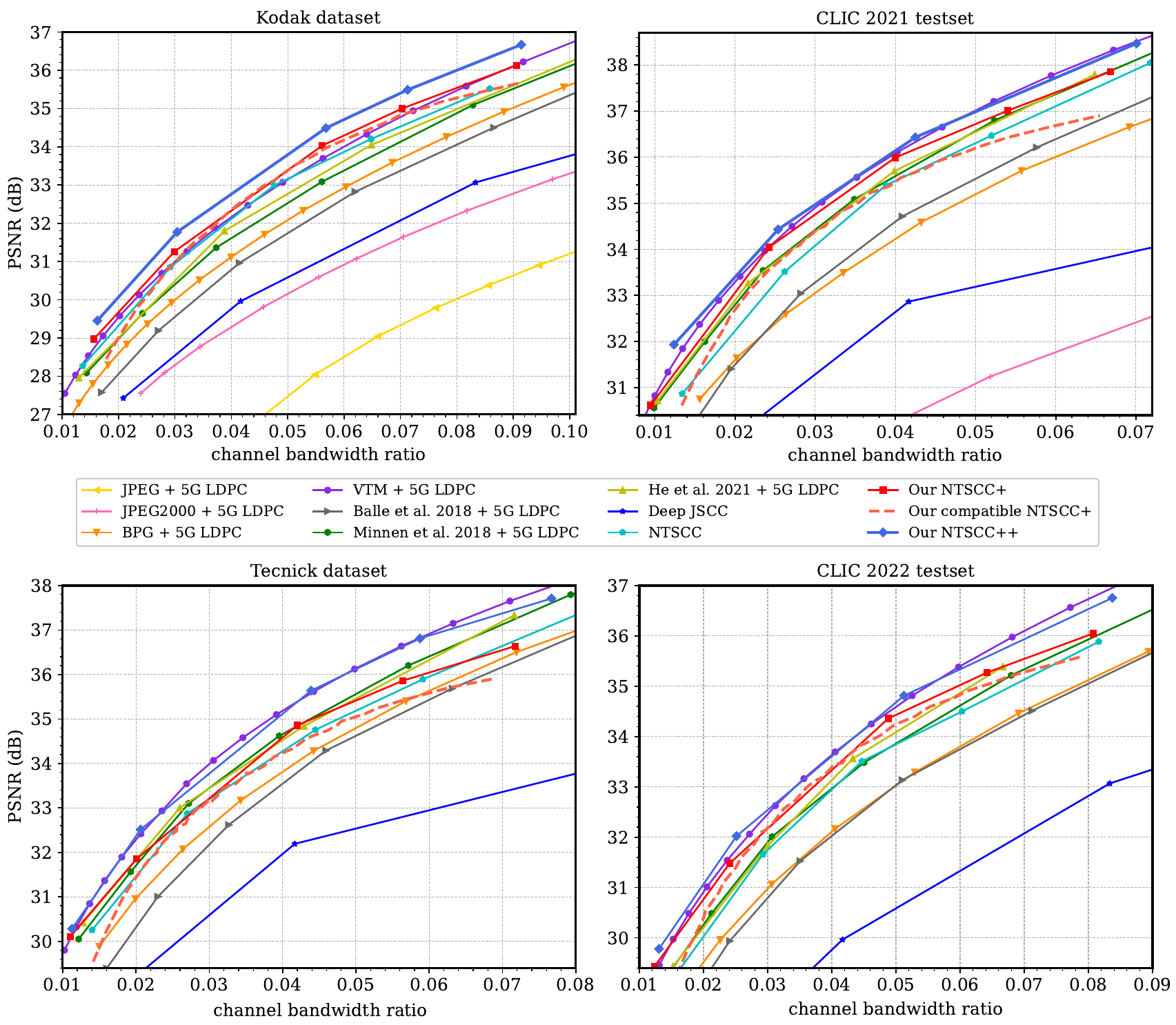}}
		\caption{End-to-end rate-distortion comparison over the AWGN channel at SNR $=10$dB for Kodak, CLIC21 testset, Tecnick, and CLIC22 testset.}
		\label{Fig_appendix_rd_curve}
		\vspace{-1.5em}
	\end{figure*}
	
	\begin{figure*}[t]
		\setlength{\abovecaptionskip}{0.cm}
		\setlength{\belowcaptionskip}{-0.cm}
		\begin{subtable}
			\centering
			\small
			\begin{tabular}{m{0.14\textwidth}<{\centering}m{0.14\textwidth}<{\centering}m{0.145\textwidth}<{\centering}m{0.145\textwidth}<{\centering}m{0.145\textwidth}<{\centering}m{0.145\textwidth}<{\centering}}
				Original image & Original patch & $\lambda_{s}=0$ & $\lambda_{s}=10^2$ & $\lambda_{s}=10^3$ &   $\lambda_{s}=10^4$
			\end{tabular}
		\end{subtable}
		\vspace{-1.5em}
		
		\begin{center}
			\hspace{-.05in}
			\subfigure[$768 \times 768$] {\includegraphics[width=0.15\textwidth]{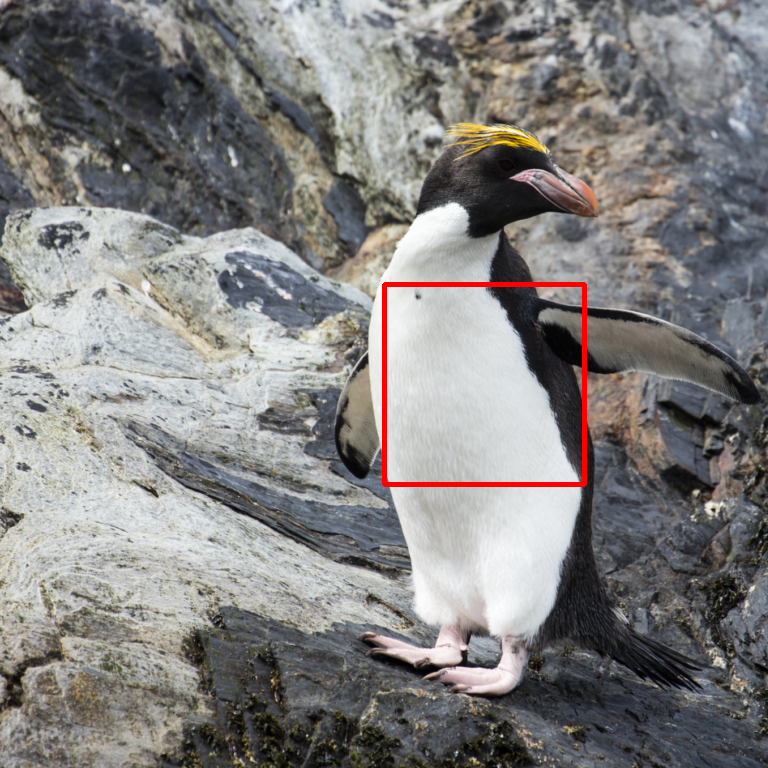}}
			\hspace{-.10in}
			\quad
			\subfigure[$\rho$ / PSNR (dB) / LPIPS] {\includegraphics[width=0.15\textwidth]{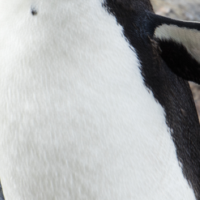}}
			\hspace{-.10in}
			\quad
			\subfigure[0.022 / 29.73 / \textbf{0.23}] {\includegraphics[width=0.15\textwidth]{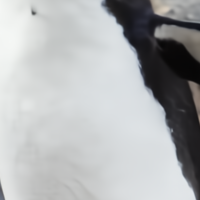}}
			\hspace{-.10in}
			\quad
			\subfigure[0.022 / 29.70 / \textbf{0.21}]{\includegraphics[width=0.15\textwidth]{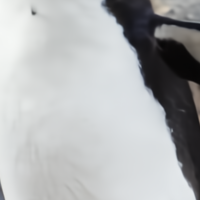}}
			\hspace{-.10in}
			\quad
			\subfigure[0.021 / 29.49 / \textbf{0.15}]{\includegraphics[width=0.15\textwidth]{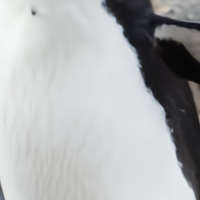}}
			\hspace{-.10in}
			\quad
			\subfigure[0.021 / 29.32 / \textbf{0.13}]{\includegraphics[width=0.15\textwidth]{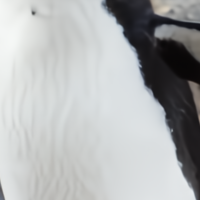}}
			
			\caption{Visual examples of online adaptation for multi-type distortions. The first and second column show the original image and its cropped patch. The third to the last column show the reconstructed images generated by NTSCC++ for different subjective loss $\lambda_{s}$ over the AWGN channel at $\text{SNR} = 10$dB. }
			\label{Fig_visual_multiloss}
		\end{center}
		\vspace{-1em}
	\end{figure*}
	
	\begin{figure*}[t]
		\setlength{\abovecaptionskip}{0.cm}
		\setlength{\belowcaptionskip}{-0.cm}
		\begin{subtable}
			\centering
			\small
			\begin{tabular}{m{0.13\textwidth}<{\centering}m{0.11\textwidth}<{\centering}m{0.121\textwidth}<{\centering}m{0.12\textwidth}<{\centering}m{0.11\textwidth}<{\centering}m{0.135\textwidth}<{\centering}m{0.1\textwidth}}
				Original image & ROI map & \shortstack{NTSCC+\\($\eta=0.2$)} & \shortstack{$\bm{m} = (0.75,1.5)$\\$\bm{\eta} = (0.15,0.3)$} &  \shortstack{$\bm{m} = (0.5,1.5)$\\$\bm{\eta} = (0.1,0.3)$} &  \shortstack{$\bm{m} = (0.25,1.5)$\\$\bm{\eta} = (0.05,0.3)$} &   \shortstack{$\bm{m} = (0,1.5)$\\$\bm{\eta} = (0,0.3)$}
			\end{tabular}
		\end{subtable}
		
		\vspace{-1em}
		
		\begin{center}
			\hspace{-.05in}
			\subfigure[$768 \times 768$] {\includegraphics[width=0.13\textwidth]{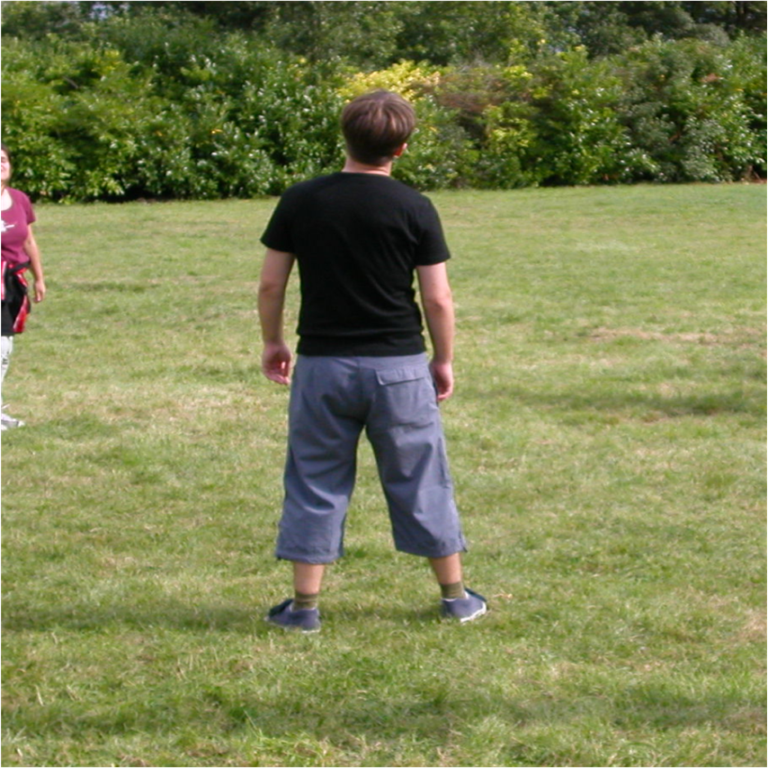}}
			\hspace{-.15in}
			\quad
			\subfigure[$\rho$/PSNR/PSNR$_{\text{ROI}}$] {\includegraphics[width=0.13\textwidth]{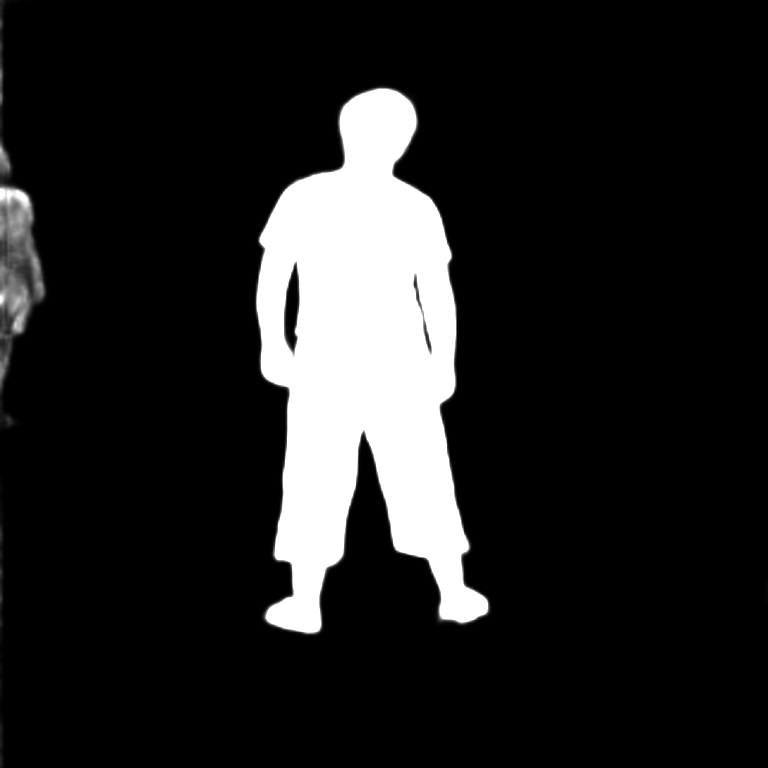}}
			\hspace{-.15in}
			\quad
			\subfigure[\textbf{0.012} / 29.8 / 31.9] {\includegraphics[width=0.13\textwidth]{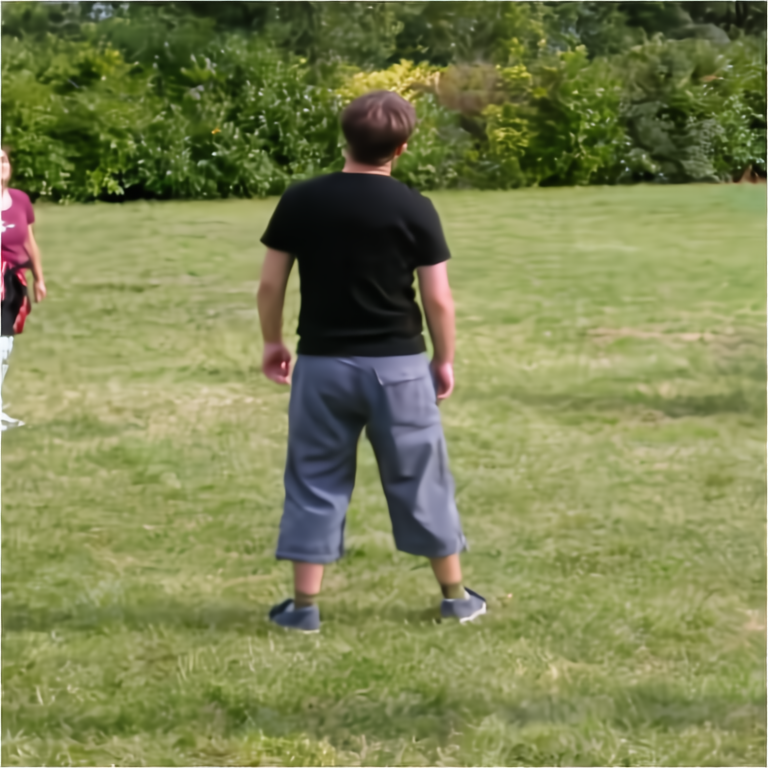}}
			\hspace{-.15in}
			\quad
			\subfigure[\textbf{0.009} / 28.6 / 31.9]{\includegraphics[width=0.13\textwidth]{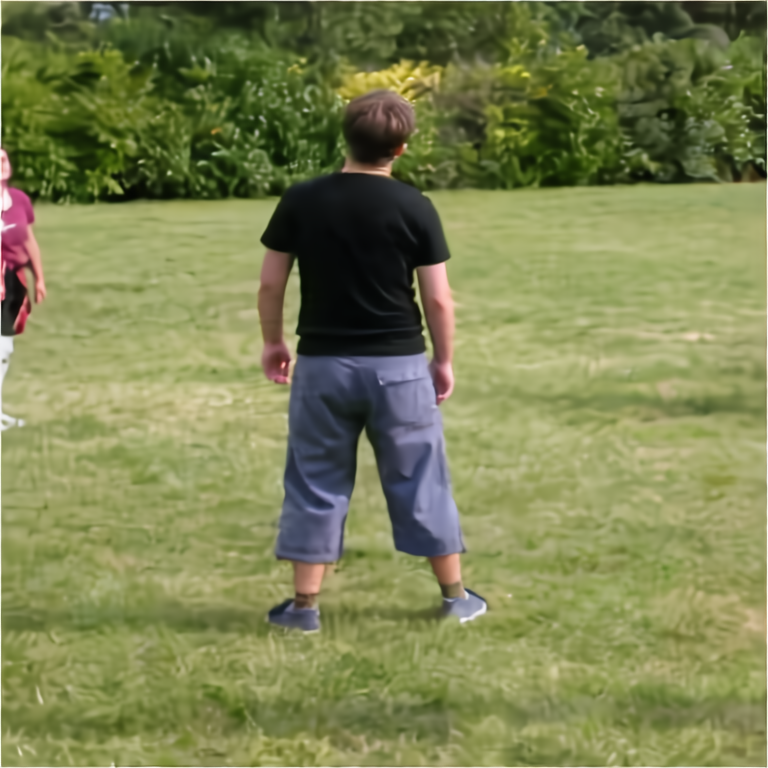}}
			\hspace{-.15in}
			\quad
			\subfigure[\textbf{0.007} / 26.7 / 32.1]{\includegraphics[width=0.13\textwidth]{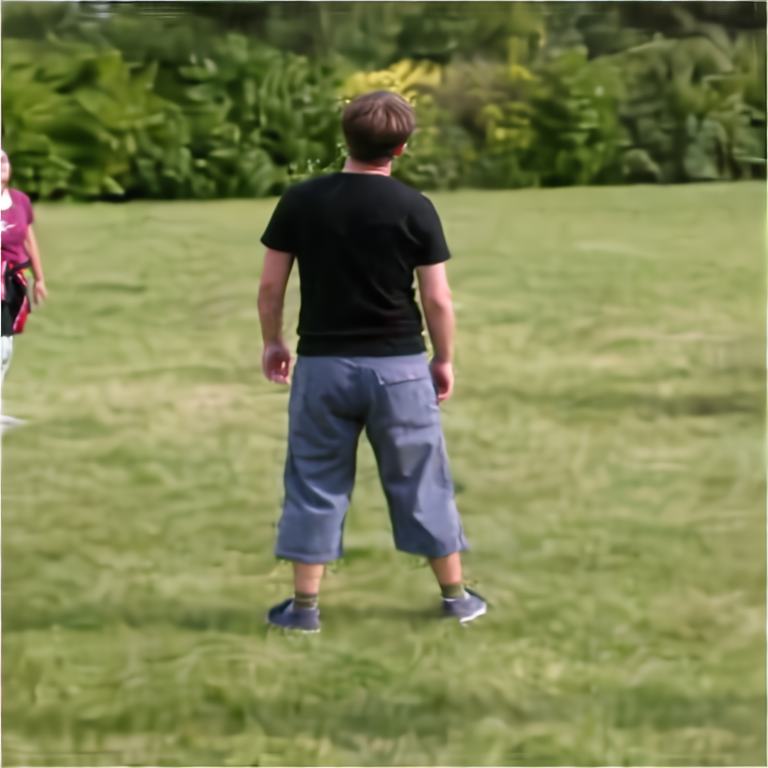}}
			\hspace{-.15in}
			\quad
			\subfigure[\textbf{0.006} / 25.1 / 32.5]{\includegraphics[width=0.13\textwidth]{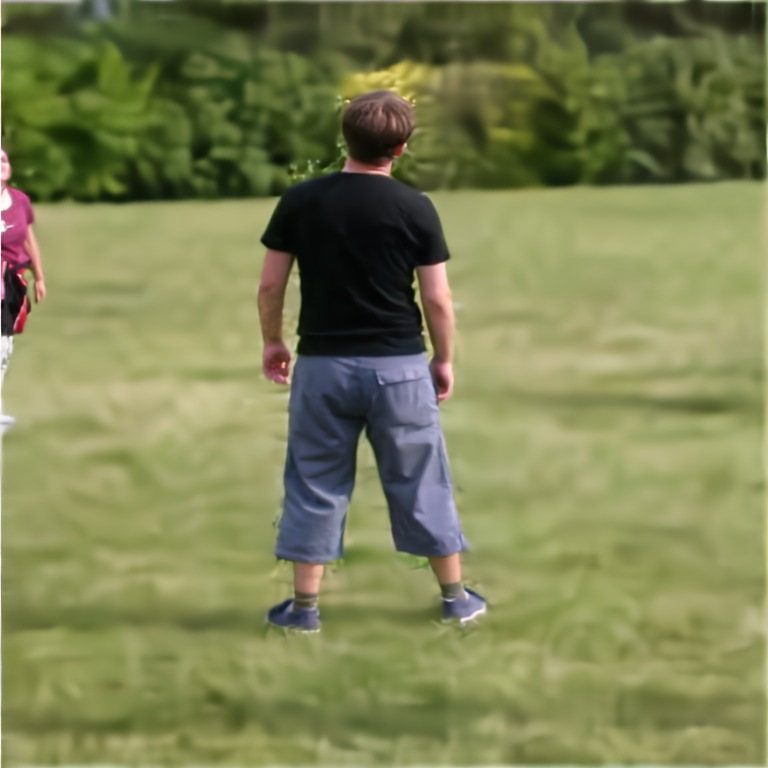}}
			\hspace{-.15in}
			\quad
			\subfigure[\textbf{0.004} / 16.5 / 32.6]{\includegraphics[width=0.13\textwidth]{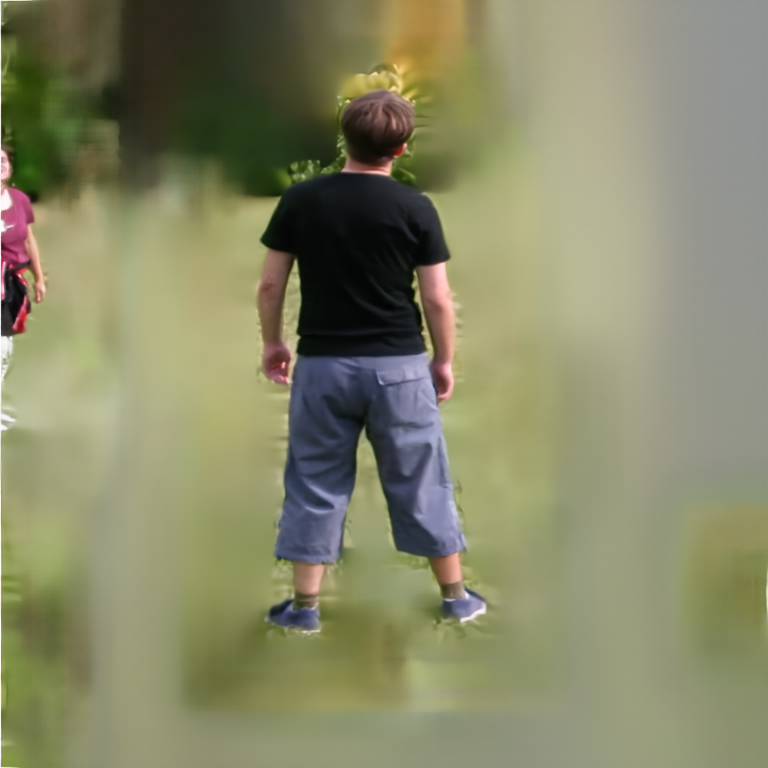}}
			
			\caption{Visual examples of online adaptation for ROI map. The first and second column show the original image and its ROI map. The third column show the reconstructed image of NTSCC+ over the AWGN channel at $\text{SNR} = 10$dB. The fourth to the seventh column show the reconstructed images by ROI-based NTSCC++. Specifically, the first number in parentheses of quality map $\bm{m}$ and scaling factor $\bm{\eta}$ represents the weight for white regions in ROI map, and the second number is the weight for black regions. We update the latents and encoder parameters for $T_{\max} = 100$ steps.}
			\label{Fig_visual_roi_2}
		\end{center}
		\vspace{-1em}
	\end{figure*}
	
	\begin{figure*}[h]
		\setlength{\abovecaptionskip}{0.cm}
		\setlength{\belowcaptionskip}{-0.cm}
		\begin{subtable}
			\centering
			\small
			\begin{tabular}{m{0.14\textwidth}<{\centering}m{0.14\textwidth}<{\centering}m{0.147\textwidth}<{\centering}m{0.147\textwidth}<{\centering}m{0.147\textwidth}<{\centering}m{0.147\textwidth}<{\centering}}
				Original image & Original patch & BPG + 5G LDPC & VTM + 5G LDPC & NTSCC+ & NTSCC++
			\end{tabular}
		\end{subtable}
		\vspace{-1.5em}
		
		\begin{center}
			\hspace{-.05in}
			\subfigure[$512 \times 768$] {\includegraphics[width=0.15\textwidth]{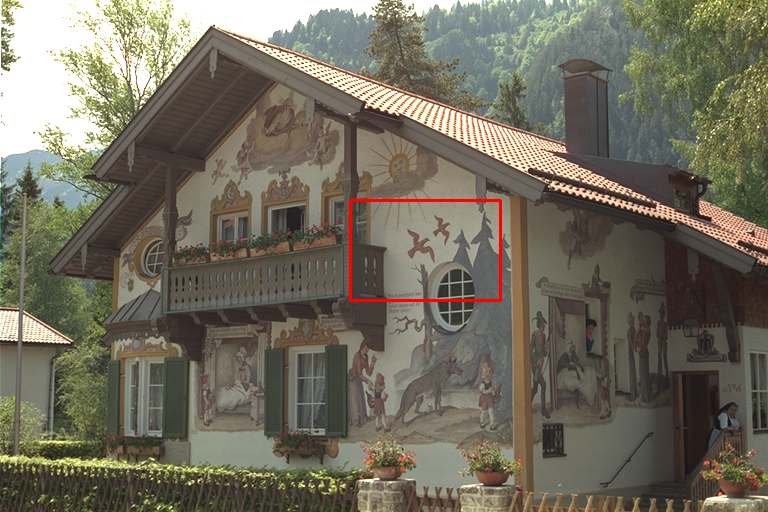}}
			\hspace{-.10in}
			\quad
			\subfigure[$\rho$ / PSNR (dB)] {\includegraphics[width=0.15\textwidth]{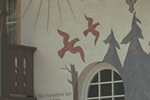}}
			\hspace{-.10in}
			\quad
			\subfigure[0.023 / 25.74] {\includegraphics[width=0.15\textwidth]{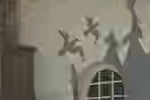}}
			\hspace{-.10in}
			\quad
			\subfigure[0.023 / 26.58]{\includegraphics[width=0.15\textwidth]{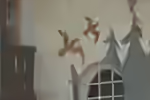}}
			\hspace{-.10in}
			\quad
			\subfigure[0.022 / 27.12]{\includegraphics[width=0.15\textwidth]{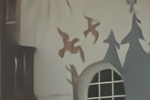}}
			\hspace{-.10in}
			\quad
			\subfigure[0.021 / 27.55]{\includegraphics[width=0.15\textwidth]{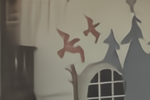}}
			
			\hspace{-.05in}
			\subfigure[$512 \times 768$] {\includegraphics[width=0.15\textwidth]{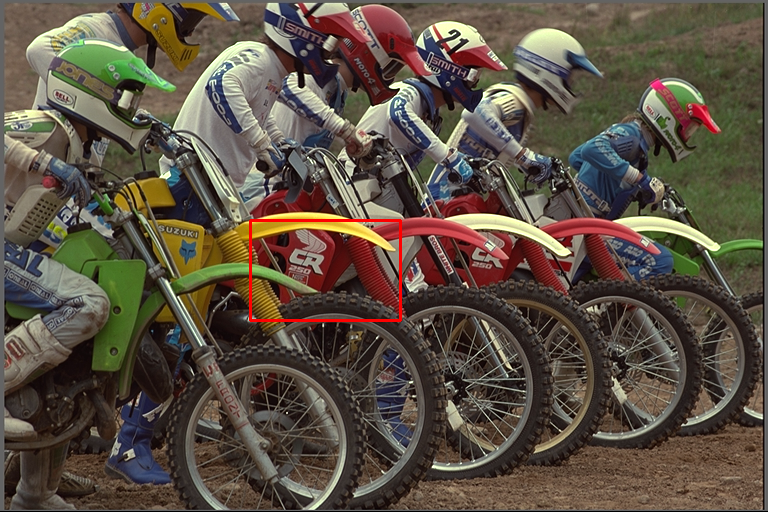}}
			\hspace{-.10in}
			\quad
			\subfigure[$\rho$ / PSNR (dB)] {\includegraphics[width=0.15\textwidth]{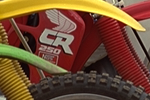}}
			\hspace{-.10in}
			\quad
			\subfigure[0.031 / 24.85] {\includegraphics[width=0.15\textwidth]{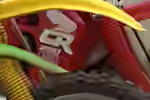}}
			\hspace{-.10in}
			\quad
			\subfigure[0.031 / 25.97]{\includegraphics[width=0.15\textwidth]{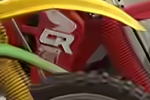}}
			\hspace{-.10in}
			\quad
			\subfigure[0.027 / 26.53]{\includegraphics[width=0.15\textwidth]{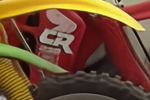}}
			\hspace{-.10in}
			\quad
			\subfigure[0.027 / 26.81]{\includegraphics[width=0.15\textwidth]{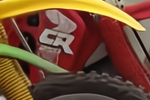}}
			
			\hspace{-.05in}
			\subfigure[$1394 \times 2048$] {\includegraphics[width=0.15\textwidth]{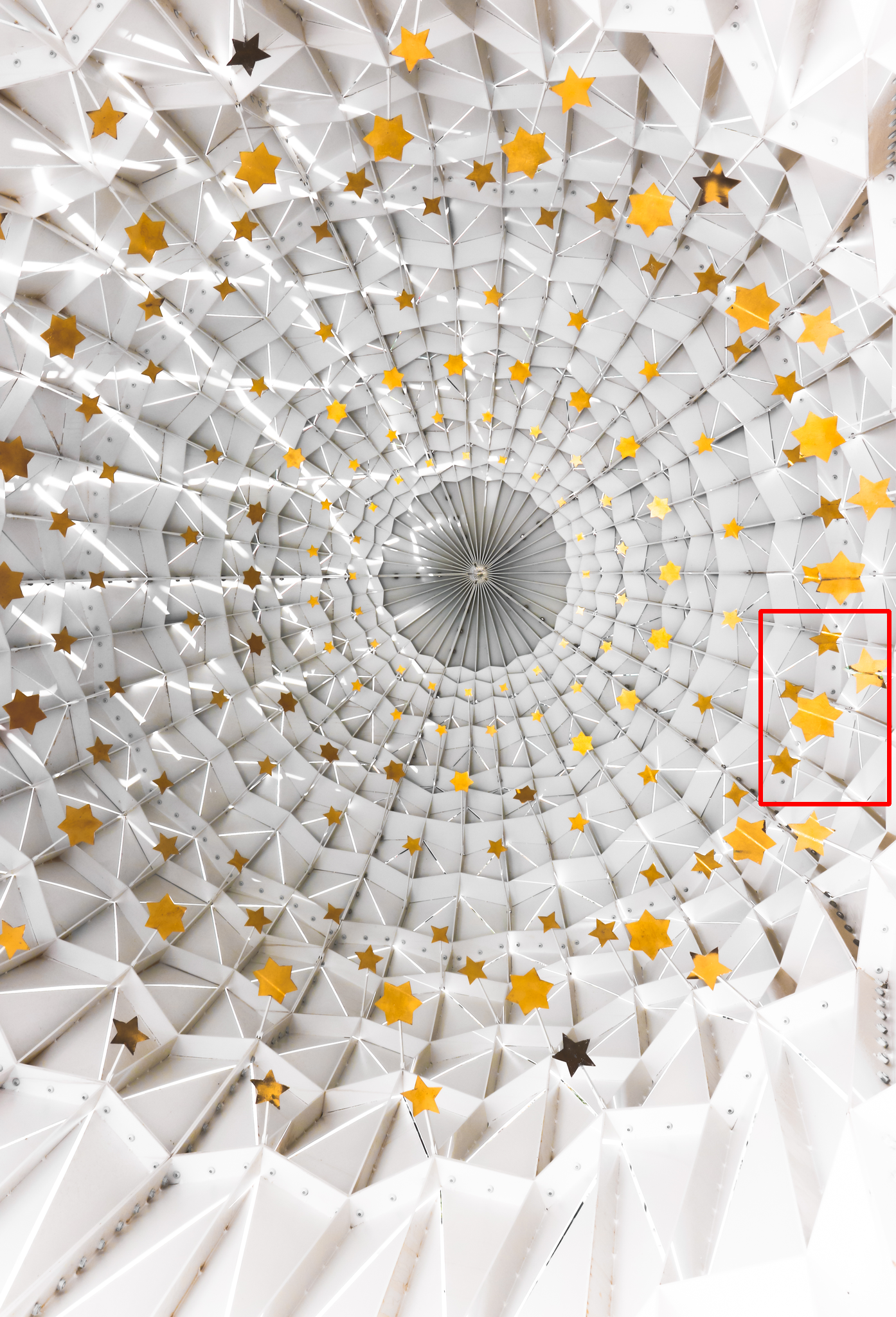}}
			\hspace{-.10in}
			\quad
			\subfigure[$\rho$ / PSNR (dB)] {\includegraphics[width=0.15\textwidth]{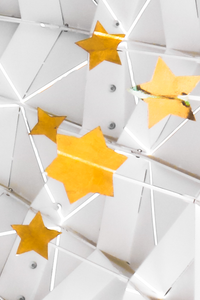}}
			\hspace{-.10in}
			\quad
			\subfigure[0.014 / 30.07] {\includegraphics[width=0.15\textwidth]{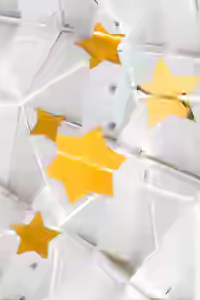}}
			\hspace{-.10in}
			\quad
			\subfigure[0.010 / 30.10]{\includegraphics[width=0.15\textwidth]{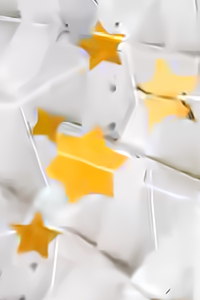}}
			\hspace{-.10in}
			\quad
			\subfigure[0.009 / 30.68]{\includegraphics[width=0.15\textwidth]{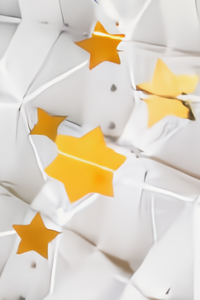}}
			\hspace{-.10in}
			\quad
			\subfigure[0.009 / 31.22]{\includegraphics[width=0.15\textwidth]{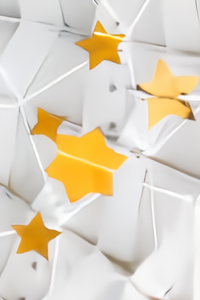}}
			
			\caption{Examples of visual comparison. The first column shows the original images. The second column shows the cropped patch in original image. The contents in third to the last columns are reconstructed by different image transmission schemes over AWGN channel at $\text{SNR} = 10$dB.}
			\label{Fig_visual_example}
		\end{center}
		\vspace{-1em}
	\end{figure*}
	
	\subsection{More Rate-Distortion Results}
	
	In Fig. \ref{Fig_appendix_rd_curve}, we offer the rate-distortion curves for an expanded set of datasets to evaluate the robustness of the proposed method in relation to the distribution of the source image. The test datasets comprise the Kodak dataset \cite{Kodak} (24 images, 512x768 pixels), CLIC 2021 test set \cite{CLIC21} (60 images, up to 2K resolution), the Tecnick dataset \cite{asuni2013testimages} (100 images, 1200x1200 pixels), and CLIC 2022 test set \cite{CLIC21} (30 images, 1365x2048 pixels). From these figures, we can see that our learned JSCC codec  consistently provide competitive transmission performance, even compared to the SOTA scheme in a wide channel bandwidth range.
	\vspace{-0.5em}
	\subsection{Visual Examples of Online Code Editing}
	
	We further show the online editing attributes of the improved NTSCC through visual examples. In Fig. \ref{Fig_visual_multiloss}, we investigate the online adaptation for multiple types of distortions, utilizing weighted MSE and LPIPS distortion as an example. It is obvious that as the weight of the subjective loss $\lambda_{s}$ increases, the reconstructions become clearer and exhibit more detailed textures.	Fig. \ref{Fig_visual_roi_2} provides an additional visual demonstration of the case for ROI-based transmission. It is observable that even when the available CBR diminishes substantially (down to 1/3 of the baseline), the quality of the ROI remains unchanged. This implies that our approach can ensure the transmission quality of salient objects even at the expense of the background quality. 
	
	\subsection{More Visualization Examples}
	
	Referring to Fig. \ref{Fig_visual_example}, we compare the reconstruction results of the proposed NTSCC+ and VTM + 5G LDPC.

\end{appendices}

\end{document}